\documentclass[aps,prd,twocolumn,twoside,superscriptaddress,floatfix, nofootinbib]{revtex4-1}

\usepackage{graphicx}
\usepackage{ifthen,amsmath,amssymb, bm}
\usepackage{graphicx}
\usepackage[dvipsnames]{xcolor}
\usepackage{hyperref}
\usepackage{float}
\usepackage{aas_macros}
\usepackage{caption}
\usepackage{subcaption}
\usepackage[export]{adjustbox}
\usepackage{bookmark}

\newcommand{\vx}{\mathbf{x}}
\newcommand{\vk}{\mathbf{k}}
\newcommand{\fnl}{f_{\rm NL}}

\newcommand{\LCDM}{$\Lambda$CDM}

\newcommand{\BH}[1]{\textcolor{black}{#1}}

\begin{document}

\title{\textsc{AbacusPNG}: Modest set of simulations of local-type primordial\\ non-Gaussianity in the DESI era}

\author{Boryana Hadzhiyska}
\email{boryanah@berkeley.edu}
\affiliation{Physics Division, Lawrence Berkeley National Laboratory, Berkeley, CA 94720, USA}
\affiliation{Berkeley Center for Cosmological Physics, Department of Physics, University of California, Berkeley, CA 94720, USA}

\author{Lehman H.~Garrison}
\affiliation{Scientific Computing Core, Flatiron Institute, 162 5th Ave, New York, NY 10010, USA}


\author{Daniel J.\ Eisenstein}
\affiliation{Center for Astrophysics $|$ Harvard \& Smithsonian, 60 Garden St., Cambridge MA 02138 USA}

\author{Simone Ferraro}
\affiliation{Physics Division, Lawrence Berkeley National Laboratory, Berkeley, CA 94720, USA}
\affiliation{Berkeley Center for Cosmological Physics, Department of Physics, University of California, Berkeley, CA 94720, USA}

\begin{abstract}

A measurement of a primordial non-Gaussianity (PNG) signal through late- or early-Universe probes has the potential to transform our understanding of the physics of the primordial Universe. While large-scale structure observables in principle contain vital information, interpreting these  measurements is challenging due to poorly understood astrophysical effects. Luckily, $N$-body simulations, such as the \textsc{AbacusPNG} set presented in this study, consisting of 9 boxes, each of size $L_{\rm box} = 2~{\rm Gpc}/h$ and particle mass of $1.01 \times 10^{10} \ M_\odot/h$, provide a viable path forward. As validation, we find good agreement between the simulations and our expectations from one-loop perturbation theory (PT) and the `separate universe' method for the matter bispectrum, matter power spectrum and the halo bias parameter associated with PNG, $b_\phi$. As a science application, we investigate the link between halo assembly bias and $b_\phi$ for halo properties known to play a vital role in accurately predicting galaxy clustering: concentration, shear (environment), and accretion rate. We find a strong response for all three parameters, suggesting that the connection between $b_\phi$ and the assembly history of halos needs to be taken into account by future PNG analyses. We further perform the first study of the $b_\phi$ parameter from fits to early DESI data of the luminous red galaxy (LRG) and quasi-stellar object (QSO) samples and comment on the effect on $f_{\rm NL}$ constraints for the allowed galaxy-halo models (note that $\sigma [f_{\rm NL}] \propto \frac{\sigma [b_\phi]}{b_\phi}$). We find that the error on $f_{\rm NL}$ is \BH{15, 8, 7} 
for the LRGs at $z = 0.5$ and $z = 0.8$ and QSOs at $z = 1.4$, respectively, suggesting that a thorough understanding of galaxy assembly bias is warranted so as to perform robust high-precision analysis of local-type PNG with future surveys. Simulations publicly available at \url{https://app.globus.org/file-manager?origin_id=ffc65d7a-0bf9-11ec-90b4-41052087bc27&origin_path=%2F}.



\end{abstract}
\maketitle



\section{Introduction} 
\label{sec:intro}

The investigation of the nature of primordial density fluctuations has long been a cornerstone of modern cosmology. Among the many tantalizing prospects of understanding these fluctuations, finding non-Gaussian features imprinted on them during the earliest moments after the Big Bang would allow us to put to the test the most widely accepted paradigms about the primordial Universe and illuminate the physics of the high-energy regime \citep{Bartolo:2004if}.

The study of primordial non-Gaussianity (PNG) encompasses a diverse array of observational avenues ranging from early- to late-Universe probes, seeking to discern subtle deviations from Gaussian statistics in the structures of our Universe. Particularly interesting for galaxy surveys is the detection and characterization of 
the so-called `local-type PNG', as this type of PNG
links the small-scale with the large-scale galaxy density distribution. A number of analyses have tried to constrain the parameter characterizing the PNG amplitude, $\fnl$, both in the cosmic microwave background (CMB) \citep{Komatsu:2003iq} and in the large-scale structure \citep{2008PhRvD..77l3514D}.

The simplest single-field models of inflation predict vanishing $\fnl$ \cite{2011JCAP...11..038C,2011JCAP...05..014T,2011JCAP...04..006B,2015JCAP...10..024D}, and hence, any detection of a non-zero signal would impact our understanding of the mechanism that generated the seeds of structure formation. The current best constraints on local PNG come from the analysis of the CMB by the $Planck$ satellite, finding $\fnl = -0.9 \pm 5.1\ (1\sigma)$ \cite{2020A&A...641A...9P}. The tightest near-future constraints from large-scale structure surveys are theoretically predicted to reach $\sigma(\fnl) \approx 1$, as the 3D distribution of galaxies from future surveys contains more Fourier modes than the 2D CMB map \citep{Sailer:2021yzm, Ferraro:2022cmj}, contaminated by foregrounds and Silk-damping on small angular scales. While theoretically these improvements are significant, there are still difficulties associated with interpreting the observed signal. In particular, to uncover these subtle non-Gaussian imprints, we need to disentangle astrophysical and non-linear evolution effects from the features of the primordial Universe. 


Summary statistics such as higher-order correlation functions, are particularly sensitive to all PNG shapes and provide a powerful venue for conducting tests on our cosmological observables. In the case of local PNG, there is a strong response even in the galaxy two-point correlation function through the scale-dependent bias. Thus, this feature, which is absent in the matter field, allows us to discern the imprints of local PNG by analizing the galaxy power spectrum on large scales. Several attempts to measure the scale-dependent bias using quasars from spectroscopic galaxy surveys have already been made, and projections for near-future experiments have forecast tighter constraints than future CMB observations. However, challenges persist in translating PNG detections into specific primordial bispectrum amplitudes, which is crucial for validating or ruling out single-field inflationary models \citep{2022JCAP...01..033B, 2021JCAP...01..062M}.


In recent years, the utilization of state-of-the-art numerical simulations, both hydrodynamical and $N$-body, 
has revolutionized our ability to explore the complex interplay between primordial physics and large-scale structure \citep[e.g.,][]{2009MNRAS.396...85D,2010JCAP...10..022W,2023ApJ...943...64C,2023ApJ...943..178C,2023arXiv231110088F,2023arXiv231212405A}. These simulations provide an invaluable laboratory for studying the link between galaxy formation and primordial physics, which is crucial for disentangling the amplitude of PNG, parameterized by $\fnl$, from the galaxy bias response to long-wavelength fluctuations, parameterized by $b_\phi$, which are otherwise degenerate in standard scale-dependent bias analyses. 

Hydrodynamical simulations provide us with a full set of realistic galaxy properties, including galaxy colors, black hole mass, and stellar mass. Because they are very computationally expensive, the amount of volume that can be simulated is typically quite limited ($\lesssim$1${\rm Gpc}^3$). However, in recent years, the technique of `separate Universes' has been employed to understand the response of small-scale galaxy physics to long-wavelength fluctuations, rendering the volume limitation a lesser issue \citep[see e.g.,][]{2020JCAP...12..013B}. Nonetheless, relying on a single implementation of the physical model can also be problematic and bias our understanding of the relationship between local PNG and galaxy formation. On the other hand, $N$-body simulations typically have a much larger volume, but since they lack galaxy physics, one needs to use heuristic methods to paint galaxy properties on them. Luckily, recently there has been significant progress in developing high-fidelity models and applying them to large-volume $N$-body simulations \citep[e.g.,][]{2020MNRAS.493.5506H,2021MNRAS.502.3242X,2022MNRAS.510.3301Y,2023MNRAS.524.2489C}.

In this work, we present a new set of $N$-body simulations run with the Abacus code \citep{2021MNRAS.508..575G} that incorporates local-type PNG, in conjunction with the latest galaxy-halo models applied in the analysis of current cosmological spectroscopic surveys \citep{2023arXiv230606314Y} to study the connection between scale-dependent bias, $b_\phi$, and the intrinsic properties of galaxies and halos. 
\BH{Another relevant simulation suite that features PNG is the Quijote-PNG suite, which besides local-type incorporates two other types of PNG. While its box size is a bit more modest ($L_{\rm box} = 1 \ {\rm Gpc}/h$), the suite comprises $\sim$1000 boxes and has been used to study both the information content of the halo and the matter density field \citep{2023ApJ...943...64C,2023ApJ...943..178C}.}
In Section~\ref{sec:valid} we validate the power spectrum and bispectrum against the one-loop perturbation theory (PT) prediction \citep{2022PhRvL.129b1301C} to ensure that the theoretical prediction matches the simulations on large scales. In Section~\ref{sec:halos}, we then study the response of $b_\phi$ to relevant halo properties for galaxy formation such as environment and concentration. Finally, in Section~\ref{sec:desi}, we characterize the value of $b_\phi$ for faithful representations of the large-scale experiment DESI and make simple Fisher forecasts for the prospects of detecting $\fnl$ through scale-dependent bias of the DESI galaxy tracers. We summarize our findings in Section~\ref{sec:conc}.


\section{Simulations}
\label{sec:sims}

\subsection{Description}
\textsc{AbacusPNG} consists of 10 simulations: five variations of $\fnl$ with two realizations each. The key parameters of the simulations are summarized in Tables~\ref{tab:sims} \& \ref{tab:common_sim_params}.  In brief, the simulations span $\fnl$ = \{-100, -30, 0, 30, 100\} with $4096^3$ particles in boxes of 2\,$h^{-1}$\,Gpc, yielding a particle mass of $1\times 10^{10} h^{-1} M_\odot$.

The \textsc{AbacusPNG} simulations are effectively an extension of the \textsc{AbacusSummit} simulations\footnote{\url{abacussummit.readthedocs.io/}} \citep{10.1093/mnras/stab2484}. They share the same scheme for cosmologies and phases (\texttt{c000}, \texttt{ph000}, etc) so that the two sets can be meaningfully compared. The parameters controlling the numeric accuracy of the solution are the same; likewise, the output data model is the same, including cleaned CompaSO halo catalogs \citep{10.1093/mnras/stab2980,10.1093/mnras/stac555}), light cones, power spectra, merger trees, and particle subsamples. This shared data model means that the abacusutils interface\footnote{\url{https://github.com/abacusorg/abacusutils}} can be used to interact with both sets of simulations.

Despite their similarities, the base mass resolution of \textsc{AbacusPNG} is $4.8\times$ coarser than that of \textsc{AbacusSummit}, which is $2.1\times 10^9\,h^{-1}\,M_\odot$.  And \textsc{AbacusSummit}, while containing several variations in particle mass, does not include the particular \textsc{AbacusPNG} resolution.  Therefore, \textsc{AbacusPNG} includes two vanilla (non-PNG) \LCDM\ simulations, \texttt{Abacus\_pngbase\_c000\_ph\{000,001\}}, that can be compared to \textsc{AbacusSummit} to isolate the effects of mass resolution from the effects of PNG.

The low mass resolution (compared to \textsc{AbacusSummit}) is by design, as we expect most information on PNG in DESI to come from LRGs and QSOs \citep[see Section 2.5.1 in][]{2016arXiv161100036D},
which have higher host halo mass than ELGs, for which \textsc{AbacusSummit}'s resolution was optimized. Therefore, at fixed computational budget, we prioritized more realizations and greater volume rather than greater resolution. This is also why the final redshift is 0.3, as the great majority of DESI-targeted LRGs and QSOs are at higher redshifts. 

A very similar version of Abacus was used to run \textsc{AbacusPNG} as \textsc{AbacusSummit}, except for minor improvements to accuracy of the near-field force (better numerical stability in the accumulation of partial forces); improvements to the numerical stability of the on-the-fly 2LPT scheme (more accurate representation of displacements); and various performance optimizations for Perlmutter.  Abacus was run with the same parameters controlling the numerical accuracy (in particular, multipole order and time step parameter) as \textsc{AbacusSummit}, and therefore the limits on accuracy and convergence placed by \cite{2021MNRAS.508..575G} and \cite{10.1093/mnras/stab2484} can be conservatively applied to \textsc{AbacusPNG}, too.

The output redshifts of various data products are as follows:
\begin{itemize}
    \item CompaSO halo catalogs (29 redshifts): 8.0, 5.0, 3.0, 2.75, 2.5, 2.25, 2.0, 1.85, 1.7, 1.625, 1.55, 1.475, 1.4, 1.325, 1.25, 1.175, 1.1, 1.025, 0.95, 0.875, 0.8, 0.725, 0.65, 0.575, 0.5, 0.45, 0.4, 0.35, 0.3
    \item Particle subsamples, 3\% and 7\% sets (10 redshifts): 3.0, 2.5, 2.0, 1.7, 1.4, 1.1, 0.8, 0.5, 0.4, 0.3
    \item Full outputs, ``Partial'' list (7 redshifts): 3.0, 2.5, 2.0, 1.4, 0.8, 0.5, 0.3
    \item Full outputs, ``Full'' list (10 redshifts): 3.0, 2.5, 2.0, 1.7, 1.4, 1.1, 0.8, 0.5, 0.4, 0.3
    \item Light cones: shells are output at every time step, approximately 800 epochs with 3 ``observers'' (same configuration as \textsc{AbacusSummit})
\end{itemize}
The \texttt{ph000} simulations use the ``Full'' list of full outputs, and \texttt{ph001} uses the ``Partial'' list.

\begin{table*}
\begin{tabular}{l|r|l|l|l}
Name & $\fnl$ & Phase & Full Outputs\footnote{``Full Outputs'' refers to the set of output epochs for which complete particle snapshots are stored (as opposed to subsamples). See Section~\ref{sec:sims} for details.} & Comment \\[0.5ex]
\hline
\texttt{Abacus\_pngbase\_c000} & 0 & \texttt{ph000} & Full & Vanilla \LCDM\ at \texttt{pngbase} mass resolution \\
\texttt{Abacus\_pngbase\_c\{300..303\}} & \{30, -30, 100, -100\} & \texttt{ph000} & Full & $\fnl$ variations, otherwise \texttt{c000} cosmology \\
\texttt{Abacus\_pngbase\_c000} & 0 & \texttt{ph001} & Partial & Second realization, fewer full particle outputs \\
\texttt{Abacus\_pngbase\_c\{300..303\}} & \{30, -30, 100, -100\} & \texttt{ph001} & Partial & Second realization, fewer full particle outputs\\
\end{tabular}
\caption{\textsc{AbacusPNG} parameter specifications\label{tab:sims}}
\end{table*}

\begin{table}
\begin{tabular}{l|r|l}
Parameter & Value & Comment \\[0.5ex]
\hline
$N$ & $4096^3$ & Number of particles \\
$L$ & 2000 $h^{-1}$\,Mpc & Box size \\
$z_\mathrm{init}$ & 99 & Initial redshift \\
$z_\mathrm{final}$ & 0.3 & Final redshift \\
$\epsilon$ & 12.2 $h^{-1}$\,kpc & \parbox[t]{3.5cm}{Proper softening length,\\Plummer-equivalent} \\
$\eta_\mathrm{acc}$ & 0.25 & \parbox[t]{3.5cm}{Time step parameter,\\acceleration-based}
\\
$M_\mathrm{p}$ & $1\times 10^{10}\,h^{-1}\,M_\odot$ & Particle mass \\
\end{tabular}
\caption{Common simulation parameters for the \textsc{AbacusPNG} set of simulations.\label{tab:common_sim_params}}
\end{table}

\begin{figure*}
    \centering
    \includegraphics[width=0.32\textwidth]{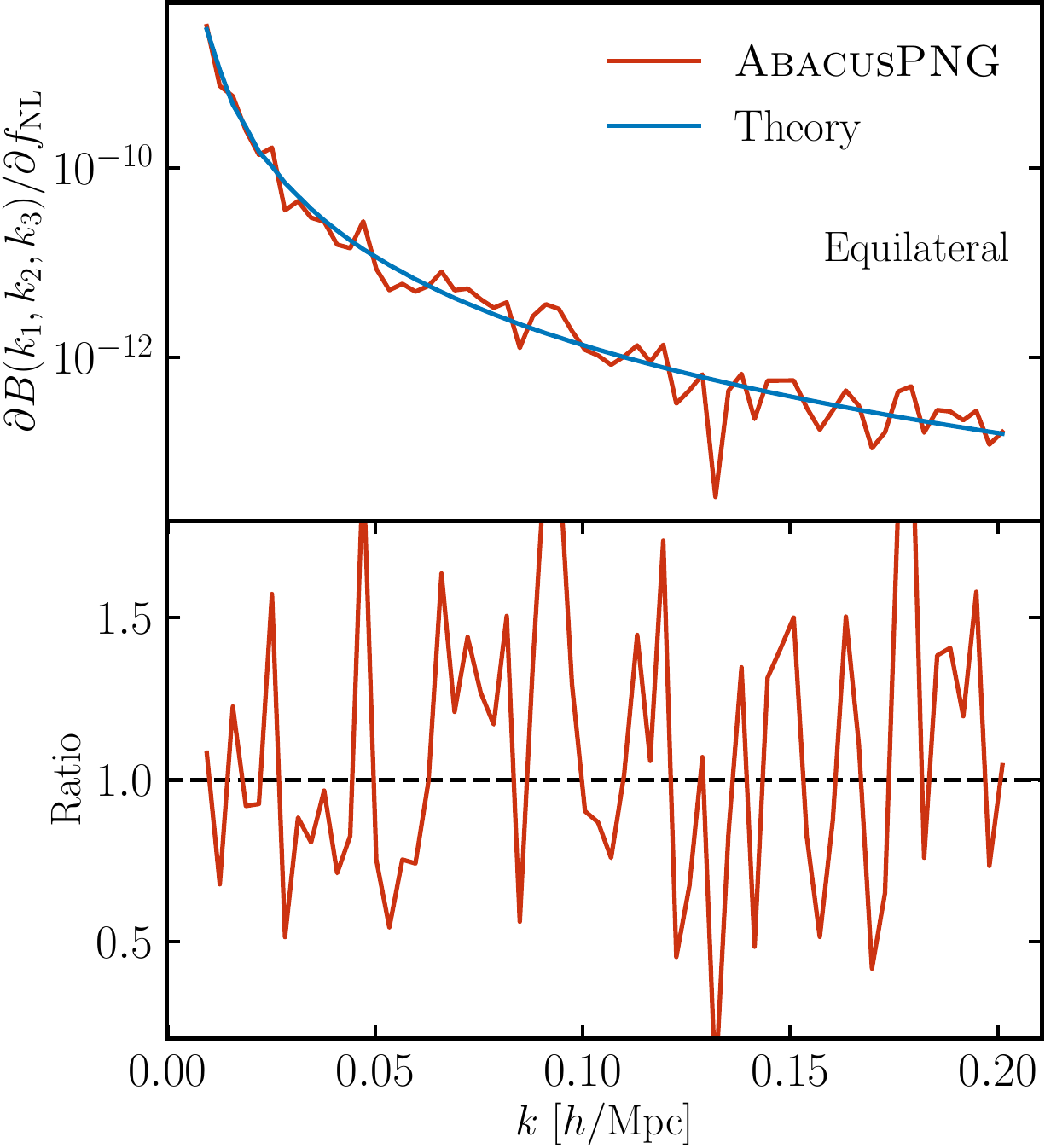}
    \includegraphics[width=0.32\textwidth]{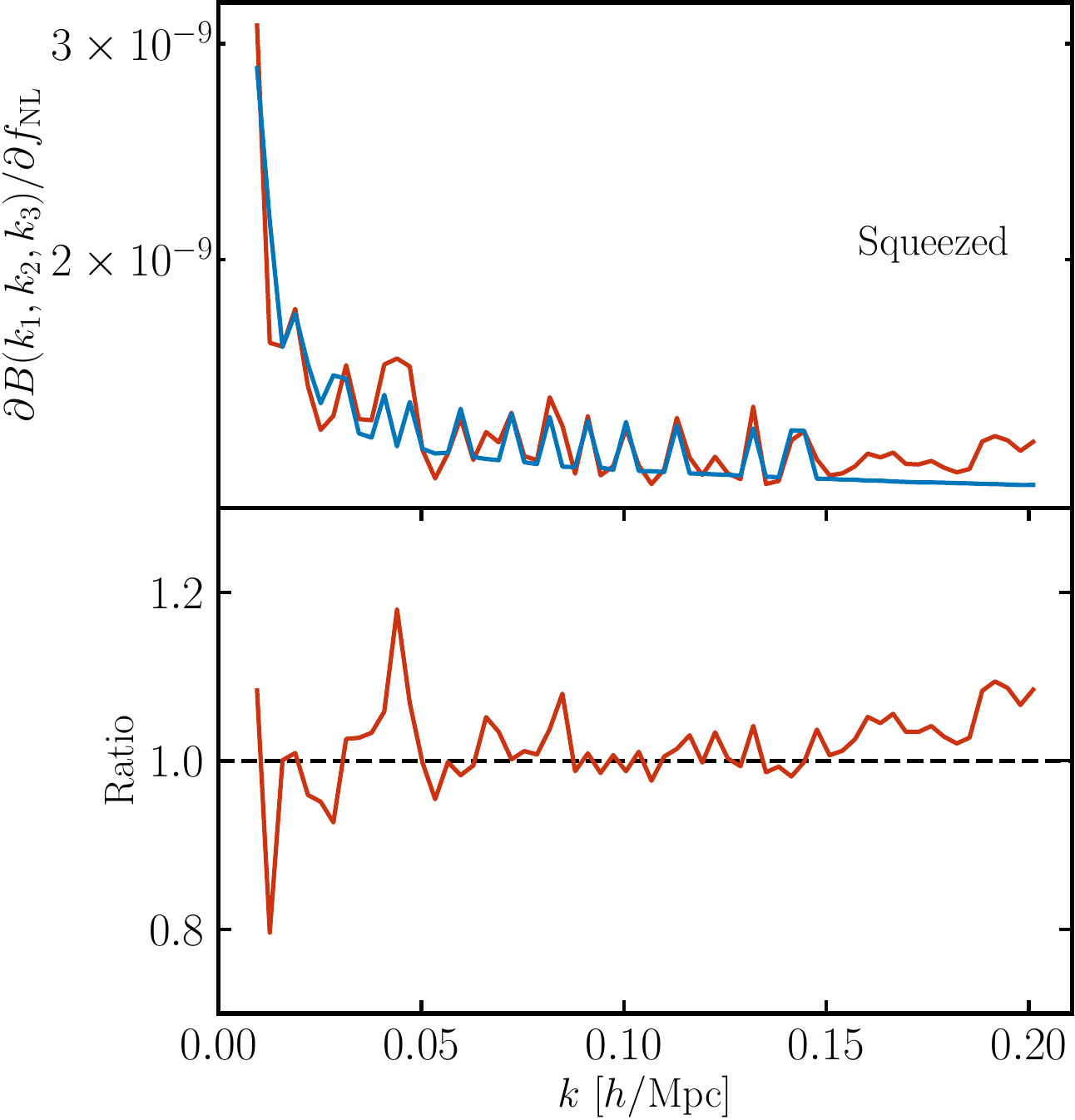}
    \includegraphics[width=0.32\textwidth]{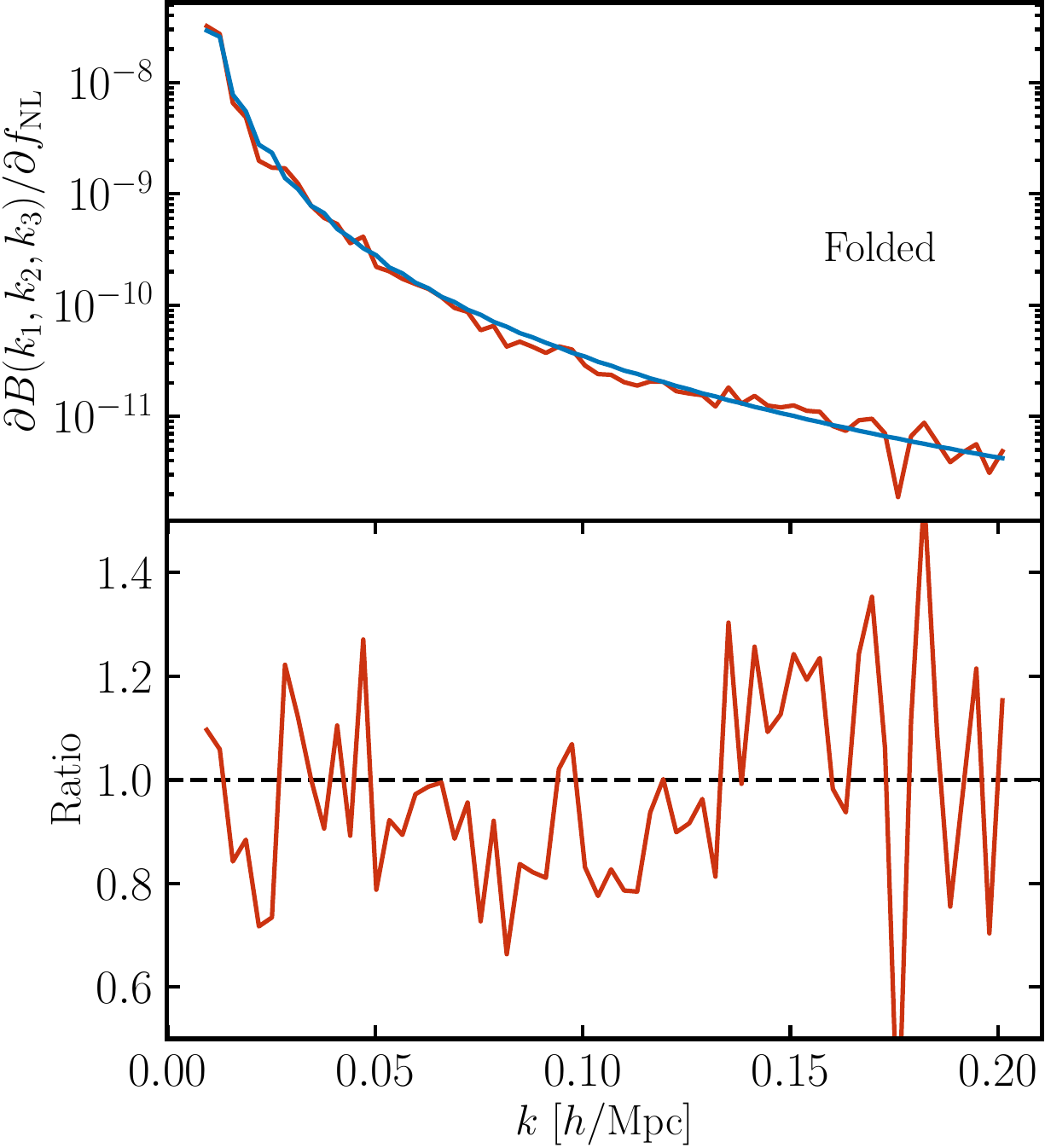}
    \caption{Validation of the effect of local-type PNG on the matter bispectrum at the initial conditions, $z_{\rm IC} = 99$. We show the derivative of the power spectrum with respect to local-type $\fnl$ as computed from the \texttt{\textsc{AbacusPNG}\_c302\_ph000} simulation as well as from theory using the tree-level approximation (see Eq.~\ref{eq:B}). We see that they are in very good agreement with each other for all three triangle configurations considered in this study: equilateral ($k_1 = k_2 = k_3$), squeezed ($k_1 = k_2 = k$, $k_3 = 3 k_{\rm F}$), folded ($k_1 = k_2 = k$, $k_3 = 2k$)). The squeezed limit yields the strongest response to local-type PNG (across all $k$-modes) and thus has the smallest error bars. For $k \gtrsim 0.15$, we see a deviation from theory, which we attribute to mild non-linearities we find in the simulation power spectrum on these scales, compared with linear theory. The noise in the theory curve is due to the fact it is computed on a grid so as to match the noise of the measurement. 
    }
    \label{fig:bk_matter}
\end{figure*}

\subsection{PNG implementation}
Local-type PNG is parameterized in terms of the primordial gravitational potential during matter domination, $\phi(\vx)$, and the parameter $\fnl$, which quantifies the amount of non-Gaussianity, via \cite{2001PhRvD..63f3002K}:
\begin{equation}
\label{eq:fnl}
\phi(\vx) = \phi_{\rm G}(\vx) + \fnl \left[\phi_{\rm G}(\vx)^2 - \left<\phi_{\rm G}(\vx)^2\right>\right],
\end{equation}
where $\phi_{\rm G}$ is a Gaussian random field.

To generate initial conditions with local-type PNG, we follow the steps outlined below:
\begin{itemize}
\item From the input power spectrum, we obtain a realization of the Gaussian primordial potential field, $\phi_{\rm G}(\vk)$.
\item We execute an inverse fast Fourier transform (FFT) to convert the field to real space, then square it and subtract the mean: $\phi_{\rm G}(\vx)^2 - \langle \phi_{\rm G}(\vx)^2 \rangle$.
\item Finally, we normalize it by the desired amplitude of local-type PNG, $\fnl$, add it back to $\phi(\vx)$, and execute a forward FFT to generate $\phi(\vk)$.
\end{itemize}
The resulting field becomes the potential source term used when running the initial conditions generator. This procedure is implemented in the zeldovich-PLT code\footnote{\url{https://github.com/abacusorg/zeldovich-PLT}} \citep{2016MNRAS.461.4125G}.

In detail, to obtain the primordial potential from the input power spectrum in the first step, the code accepts the spectral index of the primordial power spectrum. The transfer function is inferred by assuming $T(k)=1$ on large scales. This has the benefit of not requiring a separate input file, and suffices for our setup where this assumption holds true.

These initial conditions are generated at $z_\mathrm{init}=99$ using an identical procedure to \textsc{AbacusSummit}.  The initial power spectrum is that of CDM and baryons at $z=1$, backscaled to $z_\mathrm{init}$ using the linear growth factor including a non-clustering neutrino approximation.  We account for the deleterious effects of particle discreteness on the small-scale growth by applying the Particle Linear Theory rescaling of \cite{2016MNRAS.461.4125G}, with a target redshift of $z=12$. The Second-order Lagrangian Perturbation Theory (2LPT) method from that same work is applied here, via direct evaluation of the forces in the first two Abacus time steps.

\subsection{Deployment of Abacus on Perlmutter}
The following subsection is a brief report on some technical challenges and successes in running \textsc{AbacusPNG} on Perlmutter, including discussion of performance. It can be safely skipped by readers only interested in the scientific aspects of this work.

\textsc{AbacusPNG} is the first set of Abacus simulations run on NERSC's Perlmutter\footnote{\url{https://www.nersc.gov/systems/perlmutter/}}, an HPE Cray Shasta system. The simulations were run on Perlmutter's GPU nodes using DESI's compute allocation. Each simulation used 32 GPU nodes, each with 4 $\times$ NVIDIA A100 (PCIe 4.0) GPU; 1 $\times$ AMD EPYC 7763 (1 socket, 64 cores); 256 GB DDR4 RAM; and 4 $\times$ HPE Slingshot 11 NICs. We used the Cray GNU programming environment with Cray FFTW and Cray MPICH (not using any GPU-aware features or GPU-GPU communication), and CUDA 11. We will highlight some technical and performance aspects of \textsc{AbacusPNG} here but refer the reader to our more detailed reports in \cite{2019MNRAS.485.3370G} and \cite{10.1093/mnras/stab2484} for context on Abacus internals.

As in \textsc{AbacusSummit}, we used a 1D toroidal ``slab'' parallelization scheme to distribute particles across nodes. While Abacus is now capable of 2D domain decompositions, the main benefit of 2D for Abacus is to enable larger simulations where domain width per node would otherwise become too narrow. This was not required for these simulations, so we employed the 1D strategy, which has lower communication and synchronization overheads.

Abacus performed well on Perlmutter, despite far less tuning effort being devoted to it than for \textsc{AbacusSummit}. Each simulation took about 13 hours of wall clock time---about 425 node-hours. Each simulation's compute rate began around 50 M particle / sec per node at $z_\mathrm{init}$ (\texttt{singlestep} and \texttt{convolution} combined) and actually maintained that rate until the final redshift of 0.3 (about 800 time steps). This rate can be compared with 70 M particle / sec per node on Summit at the initial time, which decreased to 45 M at late times.  Fairly substantial Perlmutter overheads in launching each time step as a separate executable invocation degraded this performance to a mean of 36 M particle / sec. The overheads arose from variable performance of the NERSC file systems and Slurm job scheduler; the slowdowns were bursty and time-correlated, probably due to load on the system from unrelated jobs. Future upgrades to Abacus will avoid these overheads by running multiple time steps within a single executable invocation.

The fact that the simulations did not slow down as the particles evolved to a clustered state means that the time in the near-field computation did not exceed that of the far-field even in the simulations' most clustered state. Indeed, at the initial time the GPU overlapped 50\% of the CPU work, while at the final time it only overlapped 75\%.  The near-field work increased by $1.9\times$, but the GPU performance also increased by about $1.3\times$ (such an efficiency increase, albeit larger, was seen in \textsc{AbacusSummit} as well).

Having anticipated that, compared to \textsc{AbacusSummit}, \textsc{\textsc{AbacusPNG}}'s lower mass resolution, higher terminal redshift, and Perlmutter's large GPU-to-CPU compute ratio would result in less near-field work, we opted for a low cells-per-dimension (CPD) value of 875, shifting work from the far-field (CPU) to the near-field (GPU). At 102 particles-per-cell, this is about 50\% more particles per cell and $2.3\times$ more near-field work in the initial, unclustered state than \textsc{AbacusSummit}. Yet the $1.9\times$ growth in the near-field work was so mild that we probably could have achieved a faster time-to-solution with an even lower CPD. This mildness is attributable to the relatively large near-field/far-field transition radius in this configuration---about 5 $h^{-1}$ Mpc. Above the halo scale, the growth in the integrated 3D correlation function (that is, the growth in total number of pairs) falls off quite rapidly.

Finally, one technical hurdle to running on Perlmutter was an issue in restoring checkpoints to nodes using the shared memory scheme of \cite{2021arXiv210213140G}. The restore process simply consisted of copying files from Perlmutter Scratch to local storage using Python's \texttt{shutil} package, but such copies would hang with high frequency. The root cause was found to be an issue with the implementation of the \texttt{sendfile} Linux syscall in the Lustre filesystem used by Perlmutter Scratch. \texttt{sendfile} is designed for high-performance, in-kernel copies, and issues with it on network filesystems are not uncommon. We identified the issue during the Perlmutter acceptance testing period and reported it to NERSC, along with the workaround for \texttt{shutil}, which is to set \texttt{shutil.\_USE\_CP\_SENDFILE = False}.

\section{Validation}
\label{sec:valid}

In this section, we test the output of the \textsc{AbacusPNG} simulations against expectations of the matter and halo field distributions, using theoretical predictions from PT and heuristic methods such as the `separate universe' approach.

\subsection{Matter field}
\label{sec:matter}

To validate the simulated matter field, we examine two summary statistics: the matter power spectrum and the matter bispectrum.

\subsubsection{Bispectrum}

We start our investigation with the three-point statistic, the matter bispectrum, which we compare against the tree-level prediction at the initial redshift of the simulation ($z_{\rm IC} = 99$), when the Universe was matter-dominated. We note that at such high redshifts the tree-level prediction should perform well down to small scales. As the Universe becomes more non-linear, the tree-level approximation starts to break down, and one needs to employ a higher-order perturbative expansion \citep[see e.g.,][]{2022PhRvL.129b1301C}.

The matter bispectrum is defined as 
\begin{eqnarray}
&\langle \delta_m(\mathbf{k}_1) \delta_m(\mathbf{k}_2)  \delta_m(\mathbf{k}_3) \rangle =  \nonumber \\ & (2\pi)^3 \delta^{(3)}(\mathbf{k_1}+\mathbf{k}_2+\mathbf{k}_3) B(\mathbf{k}_1,\mathbf{k}_2,\mathbf{k}_3).
\end{eqnarray}
We estimate it using the \texttt{pylians3} package, which is based on the implementation of \citep{2020MNRAS.498.2887F}, and use linear bins between 0 and $k_\mathrm{max}=0.4~h/{\rm Mpc}$, of width $2 k_\mathrm{F}$, where $k_\mathrm{F}$ is the fundamental mode, $k_\mathrm{F} \equiv 2 \pi/L_{\rm box}$, and $L_{\rm box}$ is the box size of the simulation.

While there are other theoretically motivated shapes that are interesting to consider, here we restrict our analysis only to the local-type PNG \citep[see e.g.][for reviews of PNG]{2010AdAst2010E..72C,2022arXiv220308128A}. In that case, we can express the primordial bispectrum, $B_{\phi}(k_1,k_2,k_3)$, as
\begin{eqnarray}
    B_{\phi}(k_1,k_2,k_3) = & 2 \fnl P_\phi(k_1)P_\phi(k_2)+  \text{ 2 perm.}
\end{eqnarray}
where $P_\phi(k)$ is the primordial power spectrum and $\fnl$ is the amplitude of non-Gaussianity, associated with this shape. Taking into account the relationship between matter density and the primordial (Bardeen) potential:
\begin{equation}
    \delta_m(\vk, z) = \mathcal{M}(k,z)\phi(\vk),
\end{equation} 
where 
\begin{eqnarray}
\label{eq:M}
\mathcal{M}(k,z) = \frac{2}{3}\frac{k^2T_m(k)}{\Omega_{m}H_0^2}D(z),
\end{eqnarray}
$T_m$ is the matter transfer function, $D$ is the growth rate normalized to be equal to the scale factor, $a$, during matter domination (i.e., CMB convention), $\Omega_{m}$ is the fractional matter density parameter today, and $H_0$ is the Hubble expansion rate today, one can convert the primordial bispectrum into the matter bispectrum via
\begin{equation}
    B(k_1,k_2,k_3) = B_{\phi}(k_1,k_2,k_3) \mathcal{M}(k_1,z) \mathcal{M}(k_2,z) \mathcal{M}(k_3,z).
    \label{eq:B}
\end{equation}

Local-type PNG is a powerful probe of inflation. In single-field slow-roll inflationary models, the amplitude of the bispectrum is $\ll \mathcal{O}(\eta,\epsilon)$, where $\epsilon$ and $\eta$ are the slow-roll parameters \citep{2003JHEP...05..013M,2004JCAP...10..006C}. Measuring a large signature of local-type PNG would thus rule out slow-roll, single-field inflation. Observable levels of local-type PNG are predicted by a number of multi-field inflationary models, such as the curvaton and modulated reheating models \citep{2002PhLB..524....5L,2004PhRvD..69b3505D}, as well as by  some non-inflationary models \citep{2010AdAst2010E..67L}. 

In comparing simulations with theory, we confine our study to the bispectrum derivative rather than the full bispectrum, so as to isolate only the non-Gaussian contribution to the full bispectrum. In particular, we compute
\begin{eqnarray}
    &\frac{\partial B(k_1,k_2,k_3)}{\partial \fnl} = 2 (P_\phi(k_1)P_\phi(k_2)+\text{ 2 perm.}) \nonumber \\
    &\mathcal{M}(k_1,z) \mathcal{M}(k_2,z) \mathcal{M}(k_3,z).
\end{eqnarray}
We note that there are several subtleties in this calculation. To reduce the noise associated with the conversion from $B$ to $B_\phi$, we obtain the Bardeen potential by dividing the matter density field $\delta_m(\vk)$ by $\mathcal{M}(\vk, z)$ evaluated on the 3D Fourier grid. In addition, when calculating the theoretical prediction for the bispectrum derivative, we first evaluate the power spectrum $P_\phi(\vk)$ on the 3D grid, then take the product of the power spectra, and finally average over the triangle configurations of interest that satisfy the Dirac delta condition, $\delta^{(3)}(\vk_1+\vk_2+\vk_3)$. This is necessary, as the theoretical measurements calls for evaluating the average of the product rather than the product of two average quantities, i.e., $\langle a b \rangle \neq \langle a \rangle \langle b \rangle$. We find that not accounting for this effect leads to 10\% differences between theory and simulations when predicting small-angle triangle shapes such as the squeezed limit bispectrum. 

In Fig.~\ref{fig:bk_matter}, we show three different shapes for the local-type PNG matter bispectrum derivative at $z_{\rm IC} = 99$, calculated by finite-differencing the \textsc{AbacusPNG} simulation \texttt{\textsc{AbacusPNG}\_c302\_ph000} ($\fnl = 100$). We find very agreement between theory using the tree-level approximation (see Eq.~\ref{eq:B}) and simulations within 5\% for all three triangle configurations considered in this study: equilateral ($k_1 = k_2 = k_3$), squeezed ($k_1 = k_2 = k$, $k_3 = 3 k_{\rm F}$, folded ($k_1 = k_2 = k$, $k_3 = 2k$)). The agreement is best for the equilateral case for which grid effects play the smallest role. In the case of local-type PNG, this is also the shape that contributes the least to the total information. 
On the other hand, we found that both the squeezed and folded shapes are affected by grid effects (see previous paragraph). Nonetheless, we find satisfactory agreement for both shapes. In fact, we have separately conducted a test against \texttt{2LPTPNG}'s implementation of local-type PNG and have found better agreement between the two codes than with the tree-level expression. As expected, the squeezed limit yields the strongest response to local-type PNG (across all $k$-modes). 
For $k \gtrsim 0.15$, we see a deviation from theory, which we attribute to mild non-linearities we find in the simulation power spectrum on these scales, compared with the linear theory output 

\subsubsection{Power spectrum}

Next, we consider the matter power spectrum, which is defined as 
\begin{equation}
\langle \delta_m(\mathbf{k}) {\delta_m}^*(\mathbf{k'}) \rangle = (2\pi)^3 \delta^{(3)}(\mathbf{k}-\mathbf{k'}) P_{mm}(k).
\end{equation}
To compute the power spectrum in the \textsc{AbacusPNG} simulations for the LRGs (QSOs), we use the \texttt{abacusutils} code with linear bins of width 0.05 $h/{\rm Mpc}$, from 0 to $k_\mathrm{max}=0.4 \ (0.32)~h/{\rm Mpc}$.

\begin{figure}
    \centering
    \includegraphics[width=0.48\textwidth]{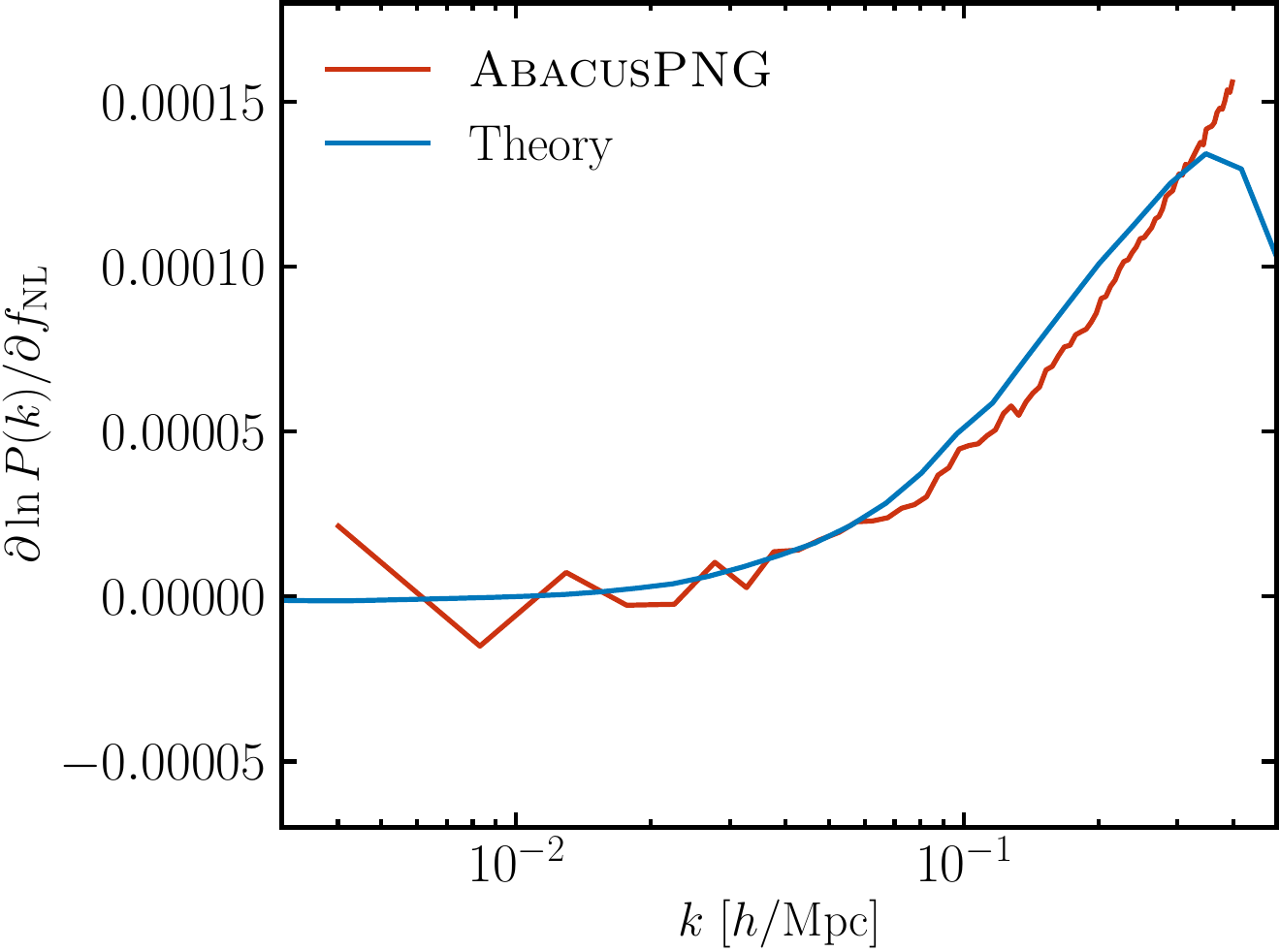}
    \caption{Validation of the effect of local-type PNG on the matter power spectrum at $z = 0.5$. We show the derivative of the power spectrum with respect to $\fnl$ as computed from the \texttt{\textsc{AbacusPNG}\_c302\_ph000} simulation as well as from theory, using 1-loop EFT (see Eq.~\ref{eq:dpdfnl}). We notice that local-type PNG has a very weak effect on the matter power spectrum -- $\sim$0.01\% at $k \approx 0.2 h/{\rm Mpc}$ The effect on large scales is negligible, whereas on small scales it grows exponentially. As expected, the agreement between theory and simulation is very good on large scales and starts to break down on non-linear scales $k \gtrsim 0.15 h/{\rm Mpc}$, where non-linearities start to become relevant.
    } 
    \label{fig:pk_matter}
\end{figure}

Since at the tree-level, the PNG contribution to the matter power vanishes (we will later see that in the case of galaxy tracers, that is not the case), we need to go to higher order in order to obtain a theoretical prediction for the matter power spectrum.
In particular, we adopt the 1-loop 
PT
expression from \citep{2022PhRvL.129b1301C}, as follows:
\begin{equation}
P(\vk) = P_{\rm G}(\vk)+\fnl P_{12}(\vk),
\end{equation}
where $P_{\rm G}$ is the standard Gaussian contribution to the power spectrum, and the PNG contribution can be calculated by solving the integral:
\begin{equation}
\label{eq:P_12}
P_{12}(\vk) = 2
\int\frac{{\rm d}^3q}{(2\pi)^3} F_2(\mathbf{q},\vk-\mathbf{q}) B(k,q,|\vk-\mathbf{q}|)
\end{equation}
where $B$ is the matter power spectrum (see Eq.~\ref{eq:B}) and $F_2$ is the standard perturbation theory (SPT) kernel defined as:
\begin{eqnarray}
    F_2(\vk_1,\vk_2) &= \frac{5}{7} + \frac{2}{7}\frac{(\vk_1\cdot \vk_2)^2}{k_1^2k_2^2} + \frac{1}{2}\frac{\vk_1\cdot \vk_2}{k_1k_2}\big(\frac{k_1}{k_2} + \frac{k_2}{k_1}\big).
\end{eqnarray}
We are only interested in the derivative with respect to $\fnl$, which becomes: 
\begin{equation}
\label{eq:dpdfnl}
    \frac{\partial P(\vk)}{\partial \fnl} = 2
\int\frac{{\rm d}^3q}{(2\pi)^3} F_2(\mathbf{q},\vk-\mathbf{q}) B(k,q,|\vk-\mathbf{q}|)
\end{equation}
To obtain the derivative, we perform Vegas Monte Carlo integration using \texttt{PyCUBA} from the \texttt{PyMultiNest} package\footnote{\url{https://github.com/JohannesBuchner/PyMultiNest}}.

Our results at $z = 0.5$ are shown in Fig.~\ref{fig:pk_matter}. We see that the theory prediction of the derivative matches the simulation result very well until $k \approx 0.1~h/{\rm Mpc}$, after which the 1-loop approximation starts to break down. As can be seen from the figure, the signature on the matter power spectrum is extremely small ($\ll$0.1\%) for all relevant scales ($k \lesssim 0.5~h/{\rm Mpc}$). This can be explained by the fact that unlike the bispectrum, $\fnl$ does not contribute at the tree-level, so one has to go to 1-loop to see an effect. On the scales where theory and simulations approximately agree, $k \approx 0.2~h/{\rm Mpc}$, the fractional change due to PNG, $\partial \ln P_{mm}/\partial \fnl$ becomes 0.0001. We attribute the remaining difference between the two to higher-order contributions. From the figure, we also see that the effect on large scales is negligible, whereas on small scales it grows exponentially. As expected, the agreement between theory and simulation is very good on large scales and starts to break down on non-linear scales $k \gtrsim 0.15 h/{\rm Mpc}$.

\subsection{Halo field}
\label{sec:halo}

\begin{figure}
    \centering
    \includegraphics[width=0.49\textwidth]{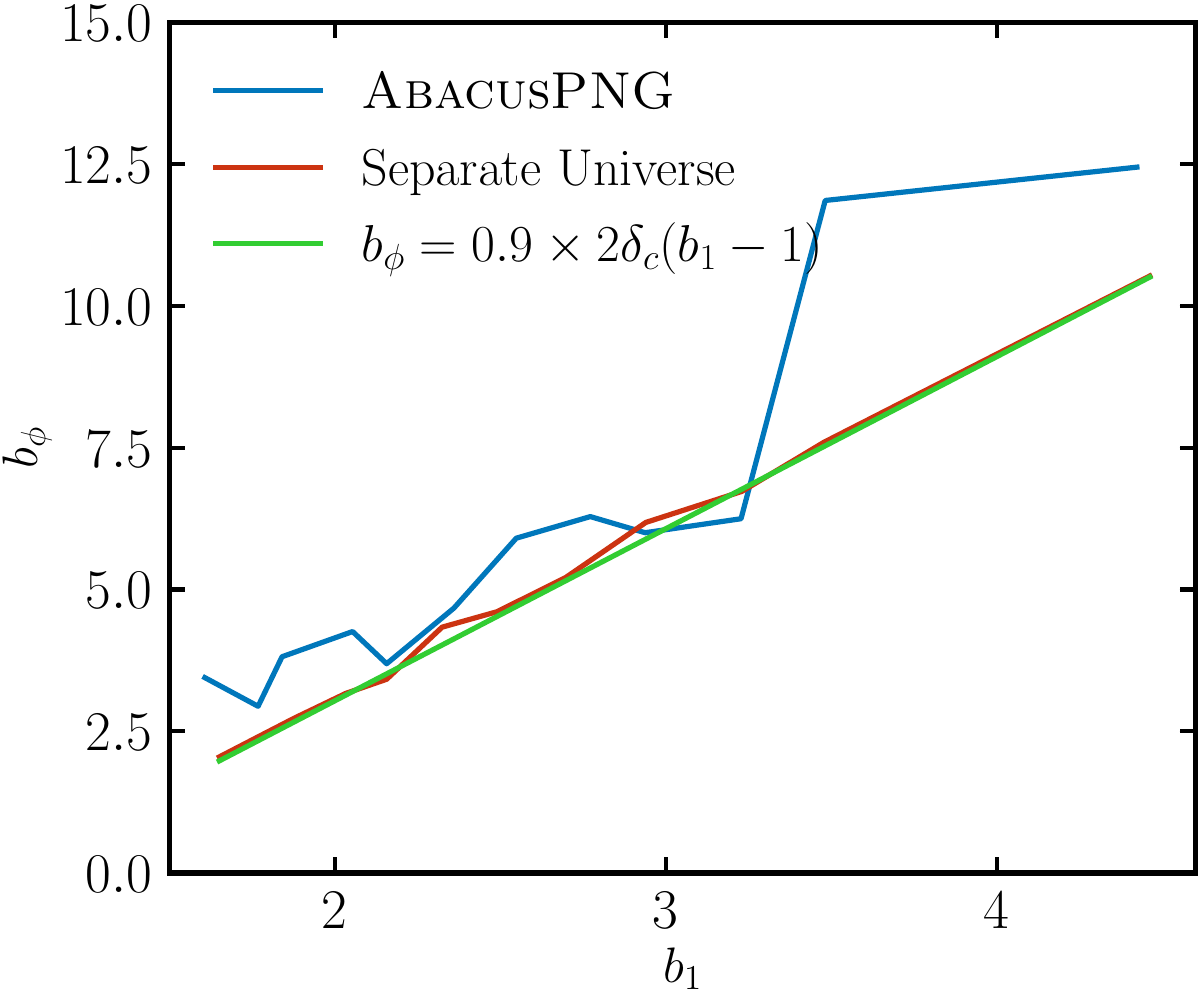}
    \caption{Comparison between $b_\phi$ obtained from the \textsc{AbacusPNG} set by fitting the power spectrum ratio (see Eq.~\ref{eq:ratio_bphi}) and $b_\phi$ obtained from the Separate Universe approach by using a pair of the \textsc{AbacusSummit} suite of simulations
    (see Eq.~\ref{eq:bphimeasure}). We note that the PNG result is obtained by fitting the ratio separately for \texttt{\textsc{AbacusPNG}\_c302\_ph000} and \texttt{\textsc{AbacusPNG}\_c302\_ph001} (with their respective $\fnl = 0$ simulations) and then averaging the two values of $b_\phi$. We also show in blue the curve coming from the modified universality relation (see Eq.~\ref{eq:bphiuniv}). We see that the halos in the Separate Universe approach are in perfect agreement with the modified universality relation. On the other hand, the \textsc{AbacusPNG} curve is substantially more noisy, as despite the fact that we have canceled most of the cosmic variance, the power spectrum ratio retains some intrinsic noise.}
    \label{fig:bphi}
\end{figure}

The distribution of halos and galaxies is a potent probe of local-type PNG via the scale-dependent bias feature in the power spectrum \cite[e.g.,][]{2008PhRvD..77l3514D,2008PhRvD..78l3519M, 2015JCAP...11..024A,2015JCAP...12..043A}. In particular, in the presence of PNG, the galaxy/halo bias expansion becomes 
\begin{eqnarray}
\label{eq:biasexp_2}
\delta_h(\vx, z)  = b_1(z)\delta_m(\vx, z) + b_\phi(z)\fnl\phi(\vx) + \epsilon(\vx),
\end{eqnarray}
where $\delta_h(\vx, z) = n_h(\vx, z)/\bar{n}_h(z) - 1$ with $n_h$ as the local number density of galaxies/halos. Similarly to the matter power spectrum, which we defined in Section~\ref{sec:matter}, we can define the galaxy/halo power spectrum as:
\begin{equation}
    \langle \delta_h(\vk) \delta_h(\vk') \rangle = (2\pi)^3 \delta^{(3)}(\vk-\vk') P_{hh}(k)  
\end{equation} 
To lowest order, we can express it as:
\begin{eqnarray}
\label{eq:Pgg}
P_{hh}(k,z) &=& b_1^2P_{mm}(k,z) + 2b_1b_\phi \fnl P_{m\phi}(k,z) \nonumber \\
&+& b_\phi^2 \fnl^2 P_{\phi\phi}(k) + P_{\epsilon\epsilon}(k) \nonumber \\
&=& \left[b_1 + \frac{b_\phi \fnl}{\mathcal{M}(k,z)}\right]^2 P_{mm}(k,z) + P_{\epsilon\epsilon},
\end{eqnarray}
where $\mathcal{M}$ is defined in Eq.~\ref{eq:M}, $P_{m \phi}$ denotes the cross-power spectrum between the primordial potential and the matter field, and $P_{\epsilon\epsilon}$ is the noise power spectrum, which we can assume to be scale-independent. Note that the cross-power term is smaller than the auto-term for $k \rightarrow 0$. We can see this by noticing that on large scales ($k \lesssim 0.01~h/{\rm Mpc}$), the presence of PNG induces scale-dependent corrections $\propto b_1b_\phi\fnl/k^2$ and $\propto b_\phi^2\fnl^2/k^4$ relative to $P_{mm}$. As we push down to larger and larger scales, the $k^{-4}$ dominates the scale-dependent signal (the transfer function is approximately equal to 1 on these scales). However, the amplitude of this effect is determined by the product $b_\phi \fnl$, rendering searches for PNG using the scale-dependent bias critically dependent on our ability to determine the bias parameter, $b_\phi$.

The bias parameter, $b_\phi$, can be estimated assuming universality of the halo mass function as \cite{2008JCAP...08..031S,2008ApJ...677L..77M,2008PhRvD..78l3507A,2010A&A...514A..46V,2012PhRvD..86f3518M,2013MNRAS.435..934F,2012PhRvD..85h3002S,2017MNRAS.468.3277B}:
\begin{eqnarray}
b_\phi(z) = 2\delta_c\left(b_1(z) - 1\right),
\end{eqnarray}
where $\delta_c = 1.686$ is the threshold overdensity for spherical collapse. Note that this is the value one obtains from linearly extrapolating to $z=0$. When comparing this relation to $N$-body simulations, recent works have found that the universality relation provides a good approximation for $b_1 \lesssim 1.5$, but overpredicts the numerical results for larger biases \citep{2009MNRAS.398..321G,2009MNRAS.396...85D,2010MNRAS.402..191P,2010JCAP...07..013R,2011PhRvD..84h3509H,2012PhRvD..85h3002S,2012JCAP...03..002W,2016JCAP...09..007B,2017MNRAS.468.3277B}. We note that this relation also depends on the definition of halo mass definition. A commonly adopted modification takes the following form:
\begin{equation}
    b_\phi(z) = 2 c \ \delta_c\left(b_1(z) - p\right),
\label{eq:bphiuniv}
\end{equation}
with $c \approx 0.9$ and $p = 1, 1.6$, where the latter value is adopted for quasars, as they are believed to be `recently accreted' \cite{2008JCAP...08..031S}, whereas the former value is typically adopted for magnitude-limited and red galaxy samples.

This $b_\phi(b_1)$ relation is employed when placing constraints and making forecasts on local-type $\fnl$ from the galaxy power spectrum \cite{2008JCAP...08..031S,2011JCAP...08..033X,2013MNRAS.428.1116R,2014PhRvD..89b3511G,2014PhRvL.113v1301L,2014MNRAS.441L..16G,2015JCAP...05..040H,2019JCAP...09..010C,2008ApJ...684L...1C,2012MNRAS.422.2854G,2014arXiv1412.4872D,2014arXiv1412.4671A,2015JCAP...01..042R,2015PhRvD..92f3525A,2015MNRAS.448.1035C,2017PhRvD..95l3513D,2017PDU....15...35R}. Since the two quantities are degenerate in the galaxy power spectrum, it is evident that uncertainties on our theoretical modeling of $b_\phi$ translate directly into uncertainties on our $\fnl$ measurements. Later in this paper, we explore how the $b_\phi(b_1)$ relation changes for halos selected by different intrinsic property and also for galaxies targeted by the ongoing cosmological survey DESI.

\subsection{Separate Universe approach}
\label{sec:sepuni}

In this section, we describe the separate universe technique, which allows us to predict the galaxy bias, $b_1$, and the local-type PNG-induced bias, $b_\phi$, by invoking the equivalence between the response of galaxy formation to long-wavelength perturbations and the response of galaxy formation to changes in the background cosmology. Assuming that the physics of galaxy formation acts on much smaller scales relative to the size of the long-wavelength perturbations, these long-wavelength perturbations act as a modified background to the process of galaxy formation on small scales \citep[this is known as the `peak-background split' argument; see e.g.,][]{1984ApJ...284L...9K,1986ApJ...304...15B}. In other words, the formation of halos and galaxies at fixed cosmology in some region of space embedded in a long-wavelength fluctuation, is equivalent to the formation of galaxies and halos in a modified cosmology at cosmic mean. This constitutes the separate universe argument.

For measuring the linear bias $b_1$, the modified cosmology needs to have a different background matter density, whereas for measuring $b_\phi$, the modified cosmology needs to have a different amplitude of the primordial scalar power spectrum, $A_s$, or equivalently, the amplitude of the linear power spectrum on the scale of $8 {\rm Mpc}/h$. In this study, we obtain the linear bias, $b_1$, without the use of the separate universe technique, by fitting $b_1$ and $A$ in the power spectrum ratio:  
\begin{equation}
\label{eq:b1}
    \frac{P_{gm}(k)}{P_{mm}(k)} = b_1+A k^2,
\end{equation}
as we include the lowest-order non-linear bias, which is proportional to $k^2$.
We can similarly obtain the bias from the ratio of the auto-power spectrum and find that the two methods are generally in agreement within 3\%. Throughout the paper, we opt to quote the value of $b_1$ coming from the cross-power spectrum, as it does not contain a shot noise contribution, which if disregarded could bias our estimate of the bias.

In the presence of local-type PNG, the amplitude of the small-scale primordial power spectrum, $P_{\phi\phi}$, gets modulated by a long-wavelength perturbation of the primordial gravitational potential, $\phi(\vx)$, which impacts the formation of structure at the scale of the small-scale perturbations. In other words, local-type PNG induces a non-vanishing bispectrum in $\phi(\vx)$, which peaks in the so-called `squeezed limit'. In the squeezed limit, two of the three legs have large values in Fourier space (i.e., short-scale), whereas the third leg has a small value of $k$ (i.e., large-scale). Physically, this means that there is a large coupling between the long-wavelength perturbations of the primordial potential, $\phi(\vx)$, with the power spectrum of two short-scale modes, $k_{\tiny \rm short}$. In other words, the primordial power spectrum at some point in space, $\vx$, can be written as:
\begin{eqnarray}
\label{eq:Pks}
P_{\phi\phi}(k_{\tiny \rm short}, z | \vx) = P_{\phi\phi}(k_{\tiny \rm short}, z) \big[1 + 4\fnl\phi(\vx)\big].
\end{eqnarray}
Thus, galaxies embedded in a long-wavelength perturbation `see' locally only a spatially uniform change to the variance of the fluctuations. They form as though in a separate universe with a modified amplitude of the primordial fluctuations \cite{2008PhRvD..77l3514D,2008JCAP...08..031S}
\begin{eqnarray}
\label{eq:tildeAs}
\tilde{A}_s = A_s\left[1  + \delta A_s\right], \ {\rm with} \ \ \ \ \ \delta A_s = 4\fnl\phi_L ,
\end{eqnarray}
where $\phi_L$ is to be treated as a constant locally and denotes the amplitude of the long-wavelength potential perturbation. 

Mathematically, the PNG-induced bias, $b_\phi$, is defined as
\begin{eqnarray}
b_\phi \equiv \frac{{\rm d} \ln n_h(z)}{{\rm d} \fnl \phi},
\end{eqnarray}
which using Eq.~\ref{eq:tildeAs}, can be expressed as:
\begin{eqnarray}
\label{eq:bphidef}
b_\phi = 4 \frac{{\rm d} \ln n_h(z)}{{\rm d}\delta A_s} = 2 \frac{{\rm d} \ln n_h(z)}{{\rm d}\delta \sigma_8}
\end{eqnarray}
Thus, we can evaluate $b_\phi$ in a separate universe with a different value of $A_s$ ($\sigma_8$). In practice, when working with cosmological simulations, this amounts to generating simulations with the same initial seed as the fiducial box, but with an input power spectrum file multiplied by $\left[1  + \delta \sigma_8^2 \right]$. \BH{In observations, making this measurement directly is extremely challenging, but recent works have suggested a possible path forward that directly extracts the PNG signal from the galaxy density field \citep{2023PhRvD.107f1301G}.}

In this work, we employ the `Linear derivative' \textsc{AbacusSummit} boxes, \texttt{base\_c112\_ph000} and \texttt{base\_c113\_ph000}, which have the same initial seed as the fiducial simulation \texttt{base\_c000\_ph000}, but a different value of $\sigma_8$: namely, $\sigma_8^{\rm high} = 1.02 \times \sigma_8^{\rm fid}$ and $\sigma_8^{\rm low} = 0.98 \times \sigma_8^{\rm fid}$, respectively.

We estimate $b_\phi$ in the separate universe approach as 
\begin{eqnarray}
\label{eq:bphimeasure}
b_\phi(z) = \frac{b_\phi^{\rm high}(z) + b_\phi^{\rm low}(z)}{2},
\end{eqnarray}
where
\begin{eqnarray}
b_\phi^{\rm high}(z) = \frac{2}{\delta \sigma_8^{\rm high}}\Big[\frac{n_h^{\rm high}(z)}{n_h^{\rm fid}(z)} - 1\Big], \\
b_\phi^{\rm low}(z) = \frac{2}{\delta \sigma_8^{\rm low}}\Big[\frac{n_h^{\rm low}(z)}{n_h^{\rm fid}(z)} - 1\Big].
\end{eqnarray}
where $n_h^{\rm fid}(z)$, $n_h^{\rm high}(z)$ and $n_h^{\rm low}(z)$ is the number density of halos (galaxies) in the fiducial, high- and low-$\sigma_8$ simulations at some redshift $z$. Note that in this work, we are interested in studying the bias $b_\phi$ in different bins of halo properties (mass, concnetration). Thus, as long as we hold fixed the mass/concentration/etc. cuts fixed for all three simulations, we can estimate $b_\phi$ in each bin of some (or more than one) halo property.

\subsection{Comparing \textsc{AbacusPNG} to separate universe}
\label{sec:valid_sepuni}

As a validation of the halo catalogs of the \textsc{AbacusPNG} simulation set, we compare the inferred $b_\phi$ using the separate universe approach to $b_\phi$ obtained by fitting the scale-dependent bias in the halo power spectrum. This is complementary to validating the matter field through the bispectrum and power spectrum, as it allows us to test whether the local-type PNG implemented into the simulation correctly couples the long wavelength mode with the short scales on which halo and galaxy formation takes place, resulting in a scale-dependent bias.

In particular, to obtain $b_\phi$ in the \textsc{AbacusPNG} simulations, we perform fits to the ratio between the halo power spectrum with $\fnl \neq 0$ and the halo power spectrum with $\fnl = 0$ for pairs of simulations with the same initial seed. That way, the cosmic variance and noise associated with the halo field cancel, and the resulting ratio yields an accurate estimate of $b_\phi$. The ratio used in the fitting of $b_\phi$ is the following:
\begin{eqnarray}
\label{eq:ratio_bphi}
\frac{P_{hh}^{\fnl=100}(k)}{P_{hh}^{\fnl=0}(k)} = \left[\frac{b_1 + \frac{b_\phi \fnl}{\mathcal{M}(k,z)}}{b_1}\right]^2,
\end{eqnarray}
We note that although this is not a feasible way of measuring $b_\phi$ that could be adopted in observations, it is useful nonetheless for the purpose of testing our simulation outputs. Here, we employ the \texttt{\textsc{AbacusPNG}\_c302\_ph000} ($\fnl = 100$) and the \texttt{\textsc{AbacusPNG}\_c000\_ph000} ($\fnl = 0$) boxes. We split the halos into 12 logarithmic mass bins ranging between $10^{12}$ and $10^{14.4}~M_\odot/h$ with the last bin encompassing all halos above $10^{14.4}~M_\odot/h$. We simultaneously fit $b_1$ and $b_\phi$ from this ratio. In contrast, for the separate universe method, we obtain $b_1$ from Eq.~\ref{eq:b1} and $b_\phi$ from Eq.~\ref{eq:bphimeasure}.

We show this comparison in Fig.~\ref{fig:bphi} by fitting the power spectrum ratio (see Eq.~\ref{eq:ratio_bphi}) with \textsc{AbacusPNG} and $b_\phi$ obtained from the Separate Universe approach by using a pair of the \textsc{AbacusSummit} suite of simulations
(see Eq.~\ref{eq:bphimeasure}). We note that the PNG result is obtained by fitting the ratio separately for \texttt{\textsc{AbacusPNG}\_c302\_ph000} and \texttt{\textsc{AbacusPNG}\_c302\_ph001} (with their respective $\fnl = 0$ simulations) and then averaging the two values of $b_\phi$. Since we only have two realizations, we would not benefit from including the $\fnl = 30$ case. The halos are split into 12 logarithmic mass bins ranging from $10^{12}$ to $10^{14.4} \ M_\odot/h$, and the biases are computed for each bin \BH{by adopting the empirical relation between halo mass and linear bias from \citep{2010ApJ...724..878T}.} We also show in blue the curve coming from the modified universality relation (see Eq.~\ref{eq:bphiuniv}). We see that the halos in the Separate Universe approach are in perfect agreement with the modified universality relation. On the other hand, the \textsc{AbacusPNG} curve is substantially more noisy, as despite the fact that we have canceled most of the cosmic variance, the power spectrum ratio retains some intrinsic noise. Some of that noise is due to higher-order contributions, which we have ignored in the ratio \citep[such as the coupling between $\delta_h$ and $\phi$ via the $b_\phi,\delta$ parameter, see][for details]{2020JCAP...12..031B}, and some of it is due to poor statistics, which affects the high mass end, where we need to compute the power spectrum for $\sim$100s of halos.

\section{Halos}
\label{sec:halos}

In this section, we study the relationship between the PNG-induced bias, $b_\phi$, and halo assembly bias properties beyond its mass (linear bias). Halo assembly bias is defined as the response of the two-point clustering of halos to the values of intrinsic halo properties at fixed halo mass. We find this a relevant and interesting question to study, as recent works show that halo assembly bias can affect the selection of galaxies in modern surveys \citep[see e.g.,][]{2018MNRAS.474.5143M}. Seeing how strong of a dependence on halo assembly bias we find in \textsc{AbacusPNG} and \textsc{AbacusSummit} would give us insight into the importance of developing models for $b_\phi$ that take into account dependencies beyond linear bias ($b_1$), or interchangeably, halo mass ($M_h$).

\subsection{Assembly bias properties}
\label{sec:hab}

We first start by introducing some of the most ubiquitously studied halo assembly bias properties: concentration, shear (environment), and accretion rate.

\subsubsection{Concentration}

The link between halo concentration and accretion history has been studied extensively in the literature \citep{1997ApJ...490..493N,2002ApJ...568...52W,2014MNRAS.441..378L,2016MNRAS.460.1214L}. It has been shown that recent merger activity induces dramatic changes in halo concentrations, and that these responses linger over a period of several dynamical times, corresponding to many Gyr \citep[see, e.g.,][]{2020MNRAS.498.4450W}. Relevant to assembly bias studies is the fact that halo concentration has a bearing on both the halo occupation distribution and the halo clustering
\citep[e.g.,][]{2001MNRAS.321..559B,2014MNRAS.441..378L,2015ApJ...799..108D,2014MNRAS.441.3359D,2018MNRAS.474.5143M}.

In this work, we adopt the following proxy for the concentration of each halo:
\begin{equation}
\label{eq:conc}
c = r_{90}/r_{25},
\end{equation}
following the recommendation of \citep{2022MNRAS.509..501H}, where $r_{90}$ and $r_{25}$ are defined as the radii, within which 90\% and 25\% of the halo particles are contained inside a sphere centered on the halo center. For more details on the halo finder and virial mass definition, we refer the reader to \citep{1998ApJ...495...80B} and \citep{2022MNRAS.509..501H}.

\subsubsection{Shear}
Our procedure for obtaining the adaptive halo shear (``shear,'' for short) manipulates the smoothed particle density field into the shear field \citep[see e.g.,][]{2023MNRAS.524.2507H}. Namely, to calculate the local ``shear'' around a halo, we first compute a dimensionless version of the tidal tensor, defined as:
\begin{equation}
    T_{ij} \equiv \partial^2 \phi_R/\partial x_i \partial x_j ,
\end{equation}
where $\phi_R$ is the dimensionless potential field calculated using Poisson's equation: $\nabla^2 \phi_R = -\rho_R/\bar{\rho}$ (the subscript $R$ corresponds to the choice of smoothing scale). We then calculate the tidal shear $q^2_R$ via:
\begin{equation}
\label{eq:shear}
 q^2_R\equiv \frac{1}{2} \big[ (\lambda_2-\lambda_1)^2+(\lambda_3-\lambda_1)^2+(\lambda_3-\lambda_2)^2\big] \, ,
\end{equation}
where $\lambda_i$ are the eigenvalues of $T_{ij}$. Physically, the ``shear'' at some particular point in space measures the amount of anisotropic pulling due to gravity at a given point in space.

\subsubsection{Accretion rate}

Previous studies of local-type PNG have surmised that the $b_\phi$ parameter is sensitive to the accretion rate of a halo, as the coupling of long- and short-wavelength perturbations is sensitive to the formation epoch of halos/galaxies. In this study, we define the accretion rate as follows:
\begin{equation}
\label{eq:gamma}
    \Gamma_{\rm dyn} (t) = \frac{\Delta \log(M)}{\Delta \log(a)} = \frac{\log[M(t)] - \log[M(t-t_{\rm dyn})]}{\log[a(t)] - \log[a(t-t_{\rm dyn})]} ,
\end{equation}
where
$M \equiv M_{\rm 200m}$ (mass within 200 times the mean density of the Universe) and the subscript ``${\rm dyn}$'' refers to the dynamical time, defined as:
\begin{equation}
    t_{\rm dyn}(z) \equiv t_{\rm cross}(z) = \frac{2 R_\Delta}{v_\Delta} = 2^{3/2} t_H(z) \left( \frac{\rho_\Delta(z)}{\rho_c(z)}\right)^{-1/2},
\end{equation}
where $t_H(z) \equiv 1/H(z)$ is the Hubble time. 

\subsection{Response to PNG}
\label{sec:bphi_hab}

\begin{figure*}
    \centering
    \includegraphics[width=0.32\textwidth]{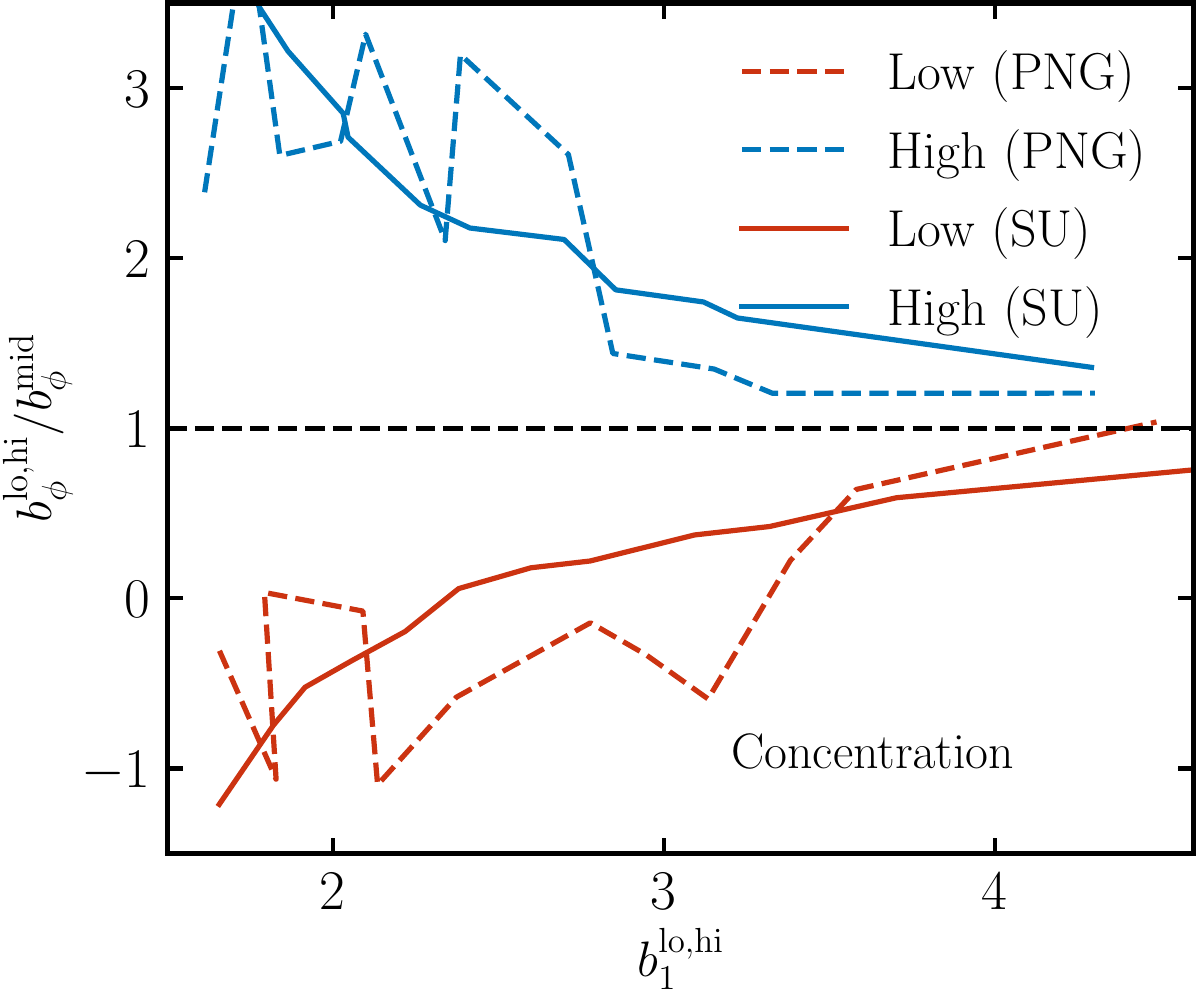}
    \includegraphics[width=0.32\textwidth]{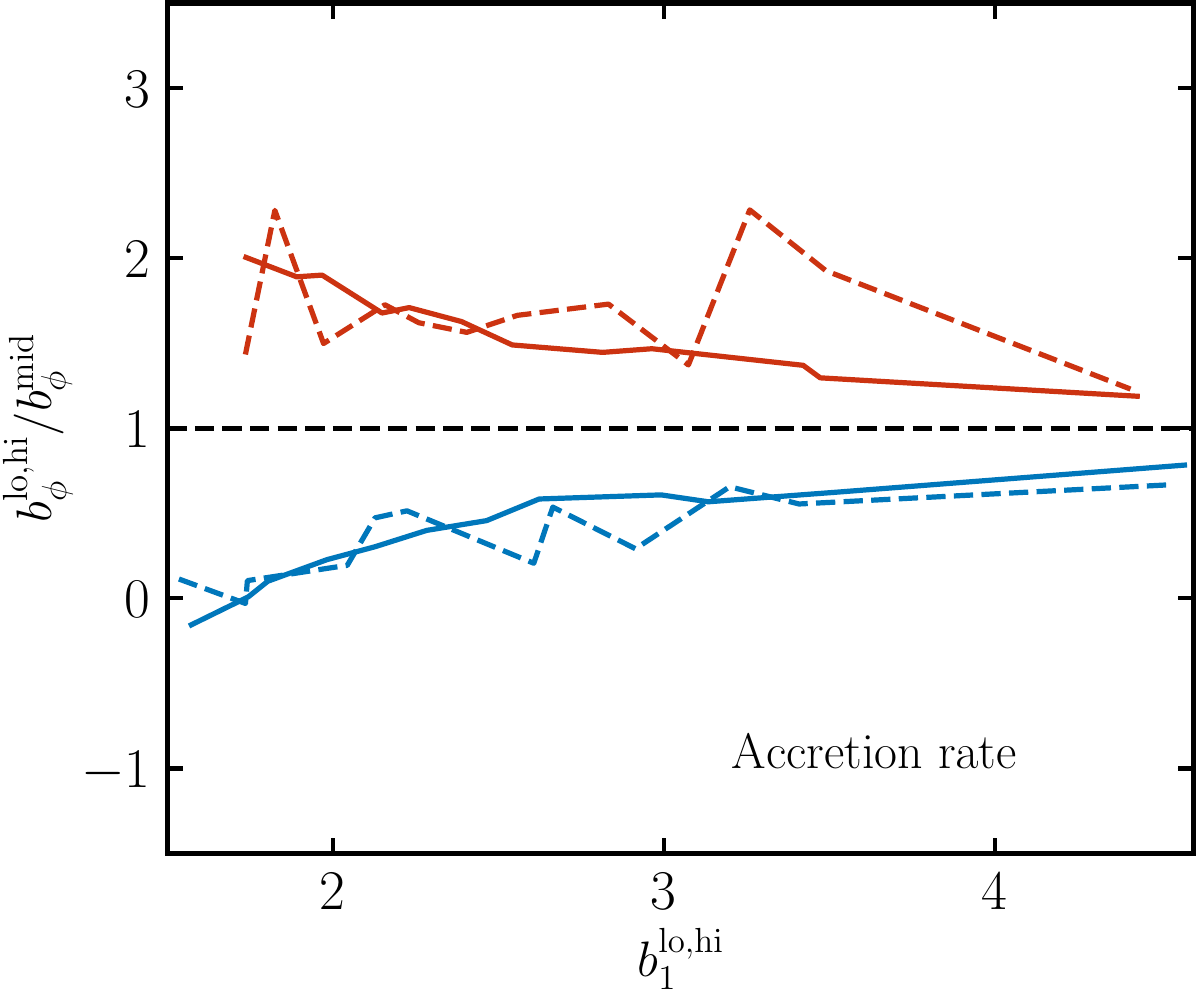}
    \includegraphics[width=0.32\textwidth]{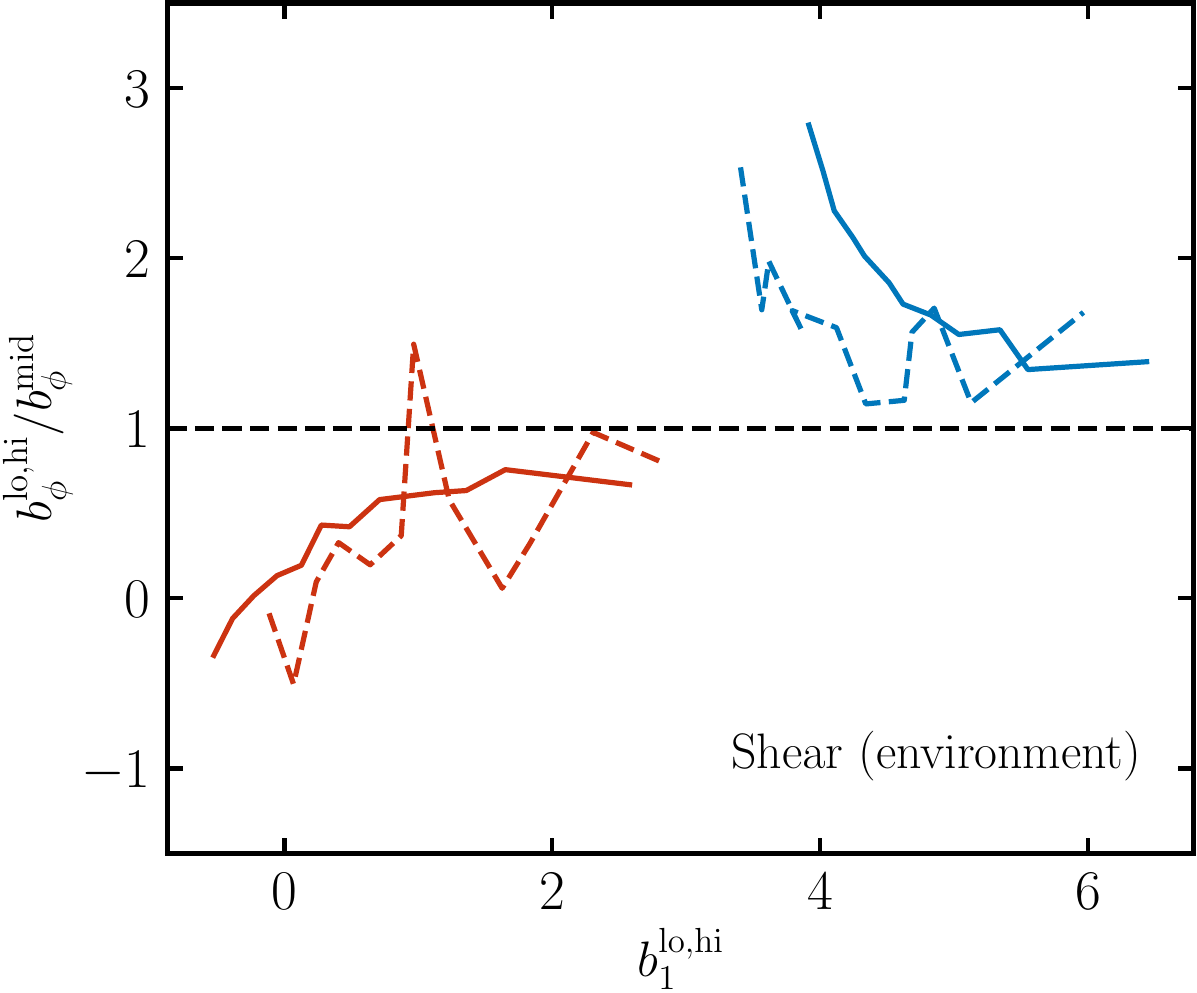}
    \caption{Response of the primordial bias parameter $b_\phi$ to the halo assembly bias properties: concentration (see Eq.~\ref{eq:conc}), accretion rate (see Eq.~\ref{eq:gamma}) and shear (see Eq.~\ref{eq:shear}). The dashed curves come from the \textsc{AbacusPNG} set of simulations, whereas the solid curves come from the Separate Universe pair of simulations, which uses the original \textsc{AbacusSummit} suite. The halos are split into 12 mass bins and then each bin is further split into 3 bins of equal sizes (33\%, 66\% percentile) based on the secondary property being considered (low, mid, high). Here, for all mass bins, we report the pairs $[b_1^{\rm high}, b_\phi^{\rm high}/b_\phi^{\rm mid}]$ in blue and $[b_1^{\rm low}, b_\phi^{\rm low}/b_\phi^{\rm mid}]$ in red. The Separate Universe and \textsc{AbacusPNG} curves are in good agreement with each other. For all three properties, the low-mass (low-bias) bins exhibit the largest variations in their values of $b_\phi$. If a given galaxy sample preferentially occupies low- or high-concentration (accretion rate) halos, then that may not reflect on the inference from the two-halo clustering, but it will affect PNG analysis. The concentration response is stronger than that of the accretion rate, and that the trend is reversed between the two, which makes sense, as actively accreting objects have lower concentration (have more spread out substructure). We note that a different definition of accretion rate might yield a stronger response. To see a change in $b_\phi$ for shear, one needs to vary the linear bias significantly at fixed halo mass.}
    \label{fig:bphi_hab}
\end{figure*}

In Fig.~\ref{fig:bphi_hab}, we show the response of the primordial bias parameter $b_\phi$ to the halo assembly bias properties: concentration (see Eq.~\ref{eq:conc}), accretion rate (see Eq.~\ref{eq:gamma}) and shear (see Eq.~\ref{eq:shear}). The dashed curves come from the \textsc{AbacusPNG} set of simulations, whereas the solid curves come from the Separate Universe pair of simulations, which uses the original \textsc{AbacusSummit} suite. The halos are split into 12 mass bins and then each bin is further split into 3 bins of equal sizes (33\%, 66\% percentile) based on the secondary property being considered (low, mid, high). Here, for all mass bins, we report the pairs $[b_1^{\rm high}, b_\phi^{\rm high}/b_\phi^{\rm mid}]$ in blue and $[b_1^{\rm low}, b_\phi^{\rm low}/b_\phi^{\rm mid}]$ in red. We note several interesting features. \BH{To convert halo mass into linear bias, we adopt the empirical relation from \citep{2010ApJ...724..878T} and calculate the linear bias of each halo before computing the statistics of interest.}

The Separate Universe and \textsc{AbacusPNG} curves are in good agreement with each other, though as noted before (see Fig.~\ref{fig:bphi}), the fitting method yields much noisier results. We note that some of the differences between the two may be due to resolution effects (which might impact the lowest- and highest-halo masses). 

For all three properties, the low-mass (low-bias) bins exhibit the largest variations in their values of $b_\phi$. In the case of concentration and accretion rate, the response of $b_\phi$ to the halo property is evidently much stronger compared with the response of $b_1$. This is crucial for modeling scale-dependent bias, as it suggests that if a given galaxy sample preferentially occupies low- or high-concentration (accretion rate) halos, then that may not reflect on the inference from the two-halo clustering (which is largely insensitive to these properties), but it will affect PNG analysis. This indicates we should aim to perform careful small-scale analysis in order to constrain the galaxy-halo connection. 

We further notice that the concentration response is stronger than that of the accretion rate, and that the trend is reversed between the two, which makes sense, as actively accreting objects have lower concentration (have more spread out substructure). On the other hand, for the case of shear, for which high-shear corresponds to high $b_\phi$, we find very different behavior: to see a change in $b_\phi$, one needs to vary the linear bias significantly at fixed halo mass. This is possible if we find that environment is indeed an important factor to consider for improving mass-only galaxy population models and resolving lensing-clustering tensions such as `Lensing is low' \citep[e.g.,][]{2021MNRAS.502.3582Y,2021MNRAS.501.1603H,2023MNRAS.521..937C}.

\section{DESI galaxies}
\label{sec:desi}

\begin{table}
    \centering
    {\renewcommand{\arraystretch}{1.5}
    \begin{tabular}{l||c|c|c}
        \hline
        \hline
        Tracer&{$\mathrm{LRG}$}&{$\mathrm{LRG}$}&{$\mathrm{QSO}$}\\[2pt]

        Redshift&
        {$0.4<z<0.6$}&{$0.6<z<0.8$}&{$0.8<z<2.1$}\\[2pt]
        
        \hline
        $\log M_\mathrm{cut}$&
        12.79$^{+0.15}_{-0.07}$&
        12.64$^{+0.17}_{-0.05}$&
        12.2$^{+0.6}_{-0.1}$\\
        
        $\log M_1$&
        13.88$^{+0.11}_{-0.11}$&
        13.71$^{+0.07}_{-0.07}$&
        14.7$^{+0.6}_{-0.6}$\\

        $\sigma$&
        0.21$^{+0.11}_{-0.10}$&
        0.09$^{+0.09}_{-0.05}$&
        0.12$^{+0.28}_{-0.06}$\\
        
        $\alpha$&
        1.07$^{+0.13}_{-0.16}$&
        1.18$^{+0.08}_{-0.13}$&
        0.8$^{+0.4}_{-0.2}$\\
        
        $\kappa$&
        1.4$^{+0.6}_{-0.5}$&
        0.6$^{+0.4}_{-0.2}$&
        0.6$^{+0.8}_{-0.2}$\\[2pt]

        \hline
        
        $\alpha_c$&
        0.33$^{+0.05}_{-0.07}$&
        0.19$^{+0.06}_{-0.09}$&
        1.54$^{+0.17}_{-0.08}$\\
        
        $\alpha_s$&
        0.80$^{+0.07}_{-0.07}$&
        0.95$^{+0.07}_{-0.06}$&
        0.6$^{+0.6}_{-0.3}$\\[2pt]

        \hline
    \end{tabular}%
    }
    \caption{Marginalized posteriors of the HOD parameters from fits to measurements of the clustering, $\xi(r_p, r_\pi)$, of DESI LRGs and QSOs in the DESI 1\% survey, performed in \citep{2023arXiv230606314Y}. For our fiducial galaxy catalogs, we adopt the best fit values of these fits at redshifts $z = 0.5$, $z = 0.8$, and $z = 1.4$ for the three samples shown here, adopting the same vanilla HOD model of AbacusHOD, and the same set of simulations, which ensures consistency of the halo mass and HOD parameter definitions. In addition, we explore extensions to the vanilla HOD model in the form of concentration and environment dependence. The error bars show $1\sigma$ uncertainties. Units of mass are given in $h^{-1}M_\odot$.
    }
    \label{tab:marg_post}
\end{table}

In this section, we explore realistic samples of DESI-like galaxies for the two tracers most relevant for PNG: luminous red galaxies (LRGs) and quasi-stellar objects (QSOs). Specifically, using halo occupation distribution (HOD) fits to early data from DESI, we investigate how $b_\phi$ responds to various extensions of the standard HOD parametrization, which are allowed by the data. We then propagate the uncertainty of measuring $b_\phi$ into an uncertainty on $\fnl$ by making simple Fisher forecasts on the combination of $b_\phi \fnl$ and report the marginalized constraints on $\fnl$ for the two tracers.

\subsection{HOD of DESI galaxies}
\label{sec:hod}

Here, we summarize the HOD model used to fit the LRG and QSO clustering with DESI 1\% data \citep[see][]{2023arXiv230606314Y}. These are the parameter values we adopt when creating our synthetic LRG and QSO catalogs. We note that we apply this HOD model (with and without extensions) to both the Separate Universe as well as the \textsc{AbacusPNG} simulations in order to estimate $b_\phi$ and derive constraints on the combination $b_\phi \fnl$.

DESI targets LRGs at $z \lesssim 1$, as they are bright galaxies with a prominent break at $4000\text{\AA}$ in their spectra, which allows them to be selected relatively easily in the data. In addition, they are highly biased tracers, which makes their BAO feature more prominent compared with other galaxy types. The DESI LRG sample has a fairly constant number density between $0.4 < z < 0.8$ of approximately $5 \times 10^{-4} [{\rm Mpc}/h]^{-3}$. QSOs (quasars), on the other hand, are the tracer choice for studying large-scale structures at high redshifts due to their extremely high luminosities. Their number density is roughly constant between $0.8 < z < 2.1$, at $2 \times 10^{-5} [{\rm Mpc}/h]^{-3}$.

The vanilla HOD model for the two samples is given by the standard formalism of \citep{2005ApJ...633..791Z}:
\begin{equation}
    \bar{n}_{\mathrm{cent}}(M) = \frac{f_\mathrm{ic}}{2}\mathrm{erfc} \left[\frac{\log_{10}(M_{\mathrm{cut}}/M)}{\sqrt{2}\sigma}\right], \label{eq:zheng_hod_cent}
\end{equation}
\begin{equation}
    \bar{n}_{\mathrm{sat}}(M) = \left[\frac{M-\kappa M_{\mathrm{cut}}}{M_1}\right]^{\alpha}
    \label{eq:zheng_hod_sat}
\end{equation}
where we note that the LRG satellite occupations equation has a modification in the form of multiplication by $\bar{n}_{\mathrm{cent}}(M)$. $M_{\mathrm{cut}}$ determines the minimum mass of a halo to host a central galaxy, $M_1$ sets the pivot scale of the power law of satellite occupation, $\sigma$ controls the steepness of the transition from 0 to 1 in the number of central galaxies, $\alpha$ is the power law index on the number of satellite galaxies, $\kappa M_\mathrm{cut}$ gives the minimum halo mass to host a satellite galaxy, $f_\mathrm{ic}$, which is a downsampling factor controlling the overall number density of the mock galaxies.

In the vanilla model, the velocity of the central galaxy is taken as the average velocity of the so-called ``L2'' subhalo \citep[see][]{2022MNRAS.509..501H}. For the satellite galaxies, the velocities are inherited from random halo particles. The analysis of \citep{2023arXiv230606314Y} also includes velocity bias, which is necessary for modeling redshift-space clustering on small scales \citep[e.g.][]{2015MNRAS.446..578G,2022MNRAS.510.3301Y}. In the AbacusHOD model, it is parametrized as: 
\begin{itemize}
    \item \texttt{$\alpha_\mathrm{vel, c}$} modulates the peculiar velocity of the central galaxy relative to the halo center along the line-of-sight (LOS):
    \begin{equation}
        v_\mathrm{cent, z} = v_\mathrm{L2, z} + \alpha_\mathrm{vel, c} \delta v(\sigma_{\mathrm{LoS}}),
        \label{eq:alpha_c}
    \end{equation}
    where $v_\mathrm{L2, z}$ denotes the LOS component of the central subhalo velocity, $\delta v(\sigma_{\mathrm{LoS}})$ denotes the Gaussian scatter, and $\alpha_\mathrm{vel, c}$ is the central velocity bias parameter.
    \item \texttt{$\alpha_\mathrm{vel, s}$} modulates how the satellite galaxy peculiar velocity deviates from that of its host particle:
    \begin{equation}
        v_\mathrm{sat, z} = v_\mathrm{L2, z} + \alpha_\mathrm{vel, s} (v_\mathrm{p, z} - v_\mathrm{L2, z}),
        \label{eq:alpha_s}
    \end{equation}
    where $v_\mathrm{p, z}$ denotes the line-of-sight component of particle velocity, and $\alpha_\mathrm{vel, s}$ is the satellite velocity bias parameter.
\end{itemize}

In Table~\ref{tab:marg_post}, we show the marginalized posteriors of the HOD parameters from fits to measurements of the clustering, $\xi(r_p, r_\pi)$, of DESI LRGs and QSOs in the DESI 1\% survey, performed in \citep{2023arXiv230606314Y}. The error bars show $1\sigma$ uncertainties. Units of mass are given in $h^{-1}M_\odot$. For our fiducial galaxy catalogs, we adopt the best fit values of these fits at redshifts $z = 0.5$, $z = 0.8$, and $z = 1.4$ for the three samples shown here, using the \textsc{AbacusSummit} boxes to ensure consistency of the halo mass and HOD parameter definitions.

\subsection{Galaxy assembly bias extensions to the model}
\label{sec:bphi}

In this analysis, we study the effect of allowing physically motivated extensions on the inferred $b_\phi$ from the Separate Universe approach. This section summarizes the assembly bias extensions allowed in the AbacusHOD model (see \cite{2022MNRAS.510.3301Y} for more details):
\begin{itemize}

    \item \texttt{$A_\mathrm{cent}$} or \texttt{$A_\mathrm{sat}$} are the concentration-based secondary bias parameters for centrals and satellites, respectively. $A_\mathrm{cent,sat} = 0$ indicate no concentration-based secondary bias. Positive values of $A$ indicate a preference for lower concentration halos, and vice versa, at fixed halo mass. The concentration definition adopted here is equivalent to the one in Section~\ref{sec:halos}.
    \item \texttt{$B_\mathrm{cent}$} or \texttt{$B_\mathrm{sat}$} are the environment-based secondary bias parameters for centrals and satellites, respectively. The environment is defined as the mass density within a $r_\mathrm{env} = 5 {\rm Mpc}/h$ tophat of the halo center, excluding the halo itself, which we note differs from the `shear' parameter we define in Section~\ref{sec:hab}. Nonetheless, we check that qualitatively these two environment parameters yield very similar results, and we choose this parametrization for consistency with the DESI analysis. $B_\mathrm{cent,sat} = 0$ indicate no environment-based secondary bias. Positive values of $B$ indicate a preference for halos in less dense environments, and vice versa, at fixed halo mass.  
\end{itemize}

In this study, we are interested in exploring how the parameter $b_\phi$ changes, as we include assembly bias extensions to the vanilla HOD model. In particular, we follow the procedure outlined below:  
    
    \BH{\textbf{We first adopt} the best-fit values from the $\xi(r_p, r_\pi)$ to the DESI 1\% data (see Table~\ref{tab:marg_post}), we construct our fiducial samples of LRGs at $z = 0.5$ and $z = 0.8$, and QSOs at $z = 1.4$. We refer to this as the fiducial or best-fit sample from hereon.}
    
    \textbf{We then create} mock catalogs in a grid centered on the fiducial parameter values. In particular, at each grid point, we vary the two mass parameters $\log M_{\rm cut}$ and $\log M_1$ by $\pm 0.5$, which roughly control the amplitude of the 2-halo and the transitioning between the 1- and 2-halo regimes, as well as the central galaxy concentration and assembly bias parameters $A_{\rm cent}$ and $B_{\rm cent}$ by $\pm 0.3$. These ranges correspond to the prior choices on these parameters in \citep{2023arXiv230606314Y}. For each parameter, we explore 5 linearly spaced values in the specified ranges (corresponding to a step size of 0.25 in the mass parameters and 0.15 in the assembly bias parameters). 
    
    \BH{\textbf{We use} mock covariance matrices generated from the 1800 small \textsc{AbacusSummit} boxes for our three samples \citep[see][]{2023arXiv230606314Y} from $\sim$2000 \textsc{AbacusSummit} small boxes by measuring the redshift-space 2D correlation function, $\xi(r_p, r_\pi)$, which can be obtained via the \citep{1993ApJ...412...64L} estimator}:
    \begin{equation}
        \xi(r_p, r_\pi) = \frac{DD - 2DR + RR}{RR},
        \label{eq:xirppi}
    \end{equation}
    where $r_p$ and $r_\pi$ are transverse and LOS separations in comoving units, respectively, whereas $DD$, $DR$, and $RR$ are the normalized data-data, data-random, and random-random pair counts in each bin of $(r_p, r_\pi)$. In principle, $\xi(r_p, r_\pi)$ captures the full information content of the two-point clustering and can thus yield stronger constraints on the galaxy-halo connection \cite{2022MNRAS.510.3301Y} compared with projected statistics such as $w_p(r_p)$, which wash out the LOS information. Here, we measure $\xi(r_p, r_\pi)$ in 
    14 logarithmically spaced bins between $\log r_{p, \mathrm{min}} = -0.83$ 
    and $\log r_{p, \mathrm{max}} = 1.5$ in the transverse direction and 
    $r_{\mathrm{\pi, max}} = 32 \ {\rm Mpc}/h$ with $\Delta r_\pi = 4 \ {\rm Mpc}/h$ 
    in the LOS direction. Thus, provided no other cuts are made (e.g., to ensure non-singularity of the covariance matrix), the number of degrees of freedom (d.o.f.) is 
    $\sim$112
    (note that we lose 4 d.o.f. due to the parameters being varied).
    We rescale the small-box (0.5$^3 \ [{\rm Gpc}/h]^3$) covariance matrices by a factor of 20, 30, and 200 for the three samples, respectively, to roughly reflect the volume of DESI Y5 for these tracers. 
    We estimate that the intrinsic uncertainty of the HOD model  in constraining the galaxy-halo connection is roughly compensated by the effective volume ratio of DESI Y5, and leave the discussion of HOD model uncertainties for future work.  

    \textbf{We compute} $\xi(r_p, r_\pi)$ for the $5^4 = 625$ grid points and using the computed mock covariance matrices, calculate the `distance' of each sample from the `true' $\xi^{\rm fid}(r_p, r_\pi)$ corresponding to the fiducial HOD sample defined in Table~\ref{tab:marg_post}:
    \begin{equation}
    \Delta \chi^2 = (\mathbf{\xi} - \mathbf{\xi}^{\rm fid})^T \mathcal{C}^{-1} (\mathbf{\xi} - \mathbf{\xi}^{\rm fid}) .
    \end{equation}

\begin{figure}
    \centering    \includegraphics[width=0.48\textwidth]{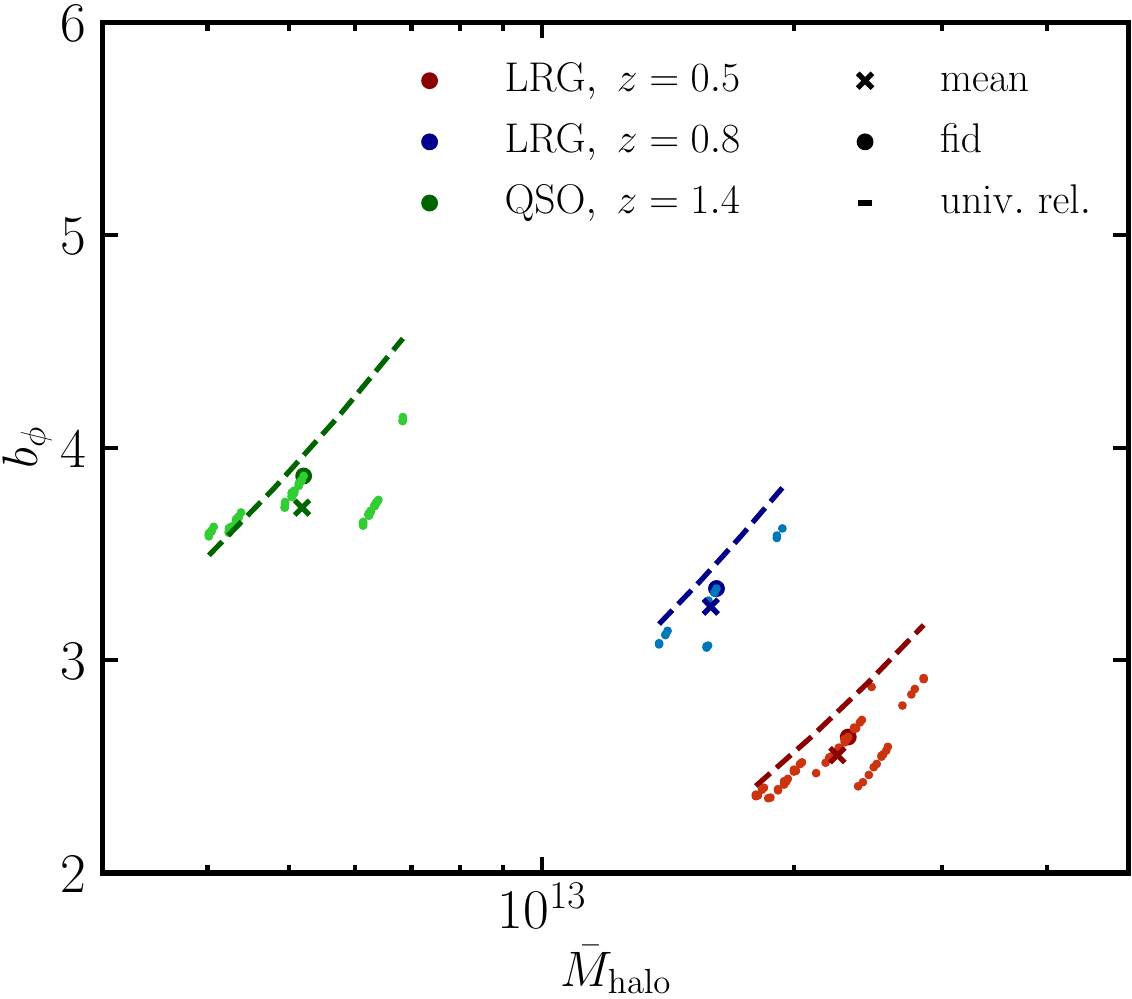}
    \caption{Scatter plot of $b_\phi$ and mean halo mass for the three samples of interest to this study: LRGs at $z = 0.5$ (red) and $z = 0.8$ (blue), and QSOs at $z = 1.4$ (green). \BH{Dots are shown for the HOD samples that satisfy the condition $\Delta \chi^2 \leq {\rm d.o.f.}$ as described in Section~\ref{sec:bphi}.} We show the mean and fiducial values for each sample with a cross and large circle, respectively, as well as the universality relation (dashed) with $c = 0.8$ (see Eq.~\ref{eq:bphiuniv}; note that a lower value of $c$ is found for the galaxies compared with the halos, $c = 0.9$). We see that the samples follow a thin slanted contour on the $b_\phi$-$\bar M_{\rm halo}$ plane. We attribute this to the fact that the mean halo mass (which is a proxy of linear bias) is relatively well constrained compared with $b_\phi$. The standard deviation of $b_\phi$ for the three samples is roughly 11\%, 6\% and 5\%, respectively.} 
    \label{fig:bphi_gal}
\end{figure}

    \textbf{Finally, we select} all samples from the 625-point grid, which have $\Delta \chi^2 \leq {\rm d.o.f.}$, 
    which corresponds to roughly $1\sigma$ deviations from the fiducial sample clustering (\BH{which is equivalent to the best-fit sample}), and calculate $b_\phi$ for each, adopting the Separate Universe approach. In Fig.~\ref{fig:bphi_gal}, we show a scatter plot of $b_\phi$ and mean halo mass for all three samples for the points that pass the $\Delta \chi^2 \leq d.o.f.$ criterion. The mean and fiducial values are shown as a cross and larger circle, respectively.
    We also show the universality relation with $c = 0.8$ (see Eq.~\ref{eq:bphiuniv}) for each sample. Intriguingly, the galaxy samples seem to prefer a lower value of $c$ than in the case of the halos (cf. $c = 0.9$). 
    We see that the samples that satisfy the condition of $\Delta \chi^2 \leq {\rm d.o.f.}$ \BH{(shown as dots on the plot)} follow a thin slanted contour on the $b_\phi$-$\bar M_{\rm halo}$ plane. We attribute this to the fact that the mean halo mass (which is a proxy of linear bias) is relatively well constrained compared with $b_\phi$. A closer look at the surviving HOD samples tells us that most of the scatter in $b_\phi$ is due to variations in the concentration parameter, $A_{\rm cent}$. Indeed, as seen in Fig.~\ref{fig:bphi_hab}, concentration has a stronger impact on $b_\phi$ than on the linear bias. It is also rather poorly constrained compared with the mass parameters due to its subtle effect on the clustering. The fiducial, mean, \BH{half difference between the maximum and minimum, and standard deviation computed using the eligible values of the parameter $b_\phi$ (i.e., from the samples that are within $1\sigma$ of the best-fit)} for the LRGs at $z = 0.5$, $z = 0.8$, and QSOs at $z = 1.4$ is 
\begin{eqnarray}
    &\{b_\phi^{\rm fid}, \ b_\phi^{\rm mean}, \ \Delta b_\phi/2, \ b_\phi^{\rm std}\} = \\ \nonumber 
    &\{2.64, 2.55, 0.29, 0.13\} \ ({\rm for \ LRG}, \ z=0.5) \\ \nonumber 
    &\{3.34, 3.25, 0.19, 0.11\} \ ({\rm for \ LRG}, \ z=0.8) \\ \nonumber 
    &\{3.87, 3.72, 0.19, 0.10\} \ ({\rm for \ QSO}, \ z=1.4)
\end{eqnarray}
This corresponds to an uncertainty of 11\%, 6\%, and 5\%, respectively. 
We additionally note that if we were able to pinpoint the mass of the tracers, the uncertainty on $b_\phi$ would decrease noticeably. While it is difficult to disentangle assembly bias and halo mass from clustering data alone, including weak lensing constraints on the galaxy-halo link would substantially help us to break HOD parameter degeneracies, allowing us to put much tighter constraints on $b_\phi$.

\textbf{Note:} \BH{We expect most of the assembly bias signal to be contained in the central galaxy population \citep[see e.g.,][]{2023MNRAS.524.2507H}. We leave the full analysis that includes running chains with all HOD extensions for later work. Beyond that, there may be other halo properties that are correlated with $b_\phi$, which may need to be further studied. It is also possible that the satellite population, which makes up between 10\% and 20\%, of all galaxies of interest to this study is sensitive to halo properties, which are degenerate with $b_\phi$, such as halo concentration and accretion rate.} We note that future data releases of the clustering of DESI LRGs and QSOs will feature much smaller error bars, which will also improve our constraints on the galaxy-halo connection and on the assembly bias of these samples.

\subsection{Fisher formalism}
\label{sec:fisher}

Here, we provide a short revision of the Fisher formalism \citep{1997ApJ...480...22T}.
For a given data vector, $\mathcal{D}$, and a vector of parameters, $\mathbf{\theta} = \{ \theta_\alpha \}$ with $\alpha$ denoting the index of each parameter, we can define the Fisher matrix as:
\begin{equation}
   F_{\alpha \beta} = \left<  \frac{\partial\log\mathcal{L}(\mathbf{D}|\mathbf{\theta})}{\partial \theta_\alpha} \frac{\partial\log\mathcal{L}(\mathbf{D}|\mathbf{\theta})}{\partial \theta_\beta} \right>,
\end{equation}
where $\mathcal{L}(D|\mathbf{\theta})$ is the likelihood. This Fisher matrix can then be used to obtain a minimum estimate of the error on each of our parameters $\theta_\alpha$ \citep{aitken_silverstone_1942}:
\begin{equation}
\sigma[\theta_\alpha] = \sqrt{(F^{-1})_{\alpha \alpha}},
\end{equation}
where $\sigma[\cdot]$ defines the error on our parameter of interest. Thus, the inverse of the Fisher information lets us infer the maximum information that we can obtain about each of our theory parameters, given some observed data. We note that $\sigma [\theta_\alpha] = F_{\alpha \alpha}^{-1/2}$ yields the non-marginalized error on the parameter $\theta_\alpha$.

In this work, we are interested in the halo and galaxy power spectrum, \BH{which we calculate using the \texttt{abacusutils} package \citep{2023OJAp....6E..38H}}. Assuming that their likelihoods are Gaussian distributed, which works particularly well for the power spectrum on the large scales we are interested in\footnote{We also note that to get an estimate of the information content, this approximation is good enough.}, we can express the log-likelihood as:
\begin{eqnarray}
   &\log\mathcal{L}(\mathbf{D}|\mathbf{\theta}) = \nonumber \\ &-\frac{1}{2} \sum_{ij}\left( P(k_i)-\bar{P}(k_i)\right)\mathcal{C}_{ij}^{-1}\left(P(k_j)-\bar{P}(k_j)\right),
\end{eqnarray} 
where $i$ and $j$ sum over all measurement bins, $P$ denotes the theory power spectrum, $\bar{P}$ is the observed power spectrum, and $\mathcal{C}$ is the covariance matrix. 

Assuming that the covariance matrix is independent of the parameters, $\theta_\alpha$, we can further simplify the Fisher matrix expression as:
\begin{equation}
    F_{\alpha\beta} = \frac{\partial\bar{P}(k_i)}{\partial\theta_\alpha}\mathcal{C}_{ij}^{-1}\frac{\partial\bar{P}(k_j)}{\partial\theta_\beta}.
\end{equation}
Thus, we only require the derivative of our summary statistic with respect to the parameters and its covariance to assess its information content. We note that in our case the covariance matrix does depend on the cosmological parameters, but neglecting this dependence gives a better approximation for the true information content \citep{2013A&A...551A..88C}. \BH{The details of how the covariance matrix is computed in this study are presented in Section~\ref{sec:bphi}}.

\subsection{Constraints on $b_\phi \fnl$}
\label{sec:bphifnl}

In this study, we are interested in assessing the information content of the scale-dependent bias and constraining $\fnl$ by robustly marginalizing over $b_\phi$.

To compute the Fisher matrix, we need a covariance matrix of the power spectrum. Here, we use an analytically calculated auto-power spectrum covariance matrices \citep{Alves2024} specifically tailored to the DESI LRG and QSO tracers at $z = 0.4 - 0.6$ and $z = 1.1 - 1.4$, respectively. We obtain the cross-power spectrum covariance matrix by 
adopting the Gaussian approximation, which should hold on large scales, and rescaling the matrix to match the Y5 volume for each tracer in each redshift range.
As validation, we compute the derivatives in two different ways: numerically (by taking finite differences) and analytically (by assuming linear theory) and find that they are in excellent agreement on large scales. We can obtain the numerical derivative with respect to 
$b_\phi \fnl$ as:
\begin{equation}
    {\frac{\partial \bar{P}(k_i)}{\partial [b_\phi \fnl]}} \approx \frac{\bar{P}(k_i)|_{[b_\phi\fnl]=100 b_\phi^\ast} - \bar{P}(k_i)|_{[b_\phi \fnl]=-100 b_\phi^\ast}}{2\delta [b_\phi \fnl]},
\end{equation}
where $\delta [b_\phi \fnl] = 100 b_\phi^\ast$ for the 
simulations used in this Section, \texttt{\textsc{AbacusPNG}\_c30\{2,3\}\_ph000}. The values of $b_\phi^\ast$ we adopt for the three samples (LRGs at $z = 0.5$, $z = 0.8$, and QSOs at $z = 1.4$) are as follows: 2.64, 3.34, 3.87. 
We compute the Fisher matrix both with and without redshift space distortions, but find that including this effect has negligible impact on the $\fnl$ constraints, as most of the information comes from large scales, which are unaffected by redshift space distortions. \BH{In redshift space, we employ the Legendre multipole expansion of the redshift-distorted power spectrum, $P(k, \mu)$:}
\begin{equation}
    P_\ell(k) = \frac{2 \ell + 1}{2} \int_{-1}^{1} P(k, \mu) L_\ell(\mu) d/\mu
\end{equation}
\BH{where $L_\ell$ is the Legendre polynomial and $P_\ell(k)$ are the multipoles of the redshift-distorted power spectrum, $P(k, \mu)$. From hereon, we drop the $\ell$ subscript to simplify the notation.} 

We also perform direct fits to the simulation measurements. For the auto-power spectrum, we vary the combination $b_\phi \fnl$, the linear bias $b_1$, and the shotnoise parameter $a$ (\BH{to allow for devations from the the assumption of Poisson shotnoise}), whereas for the cross-power spectrum case, we vary only $b_\phi \fnl$ and $b_1$.
\begin{eqnarray}
\label{eq:Pgg_Pgm}
P_{gg}(k) &=& \left[b_1 + \frac{b_\phi \fnl}{\mathcal{M}(k,z)}\right]^2 P_{mm}(k) + \frac{1}{\bar n_g}(1+a) \nonumber \\
P_{gm}(k) &=& \left[b_1 + \frac{b_\phi \fnl}{\mathcal{M}(k,z)}\right] P_{mm}(k) ,
\end{eqnarray}
where $\bar n_g$ is the mean number density of the galaxy sample and $\mathcal{M}(k,z)$ is defined in Eq.~\ref{eq:M}. The linear theory approximation adopted here starts to break down beyond $k \gtrsim 0.15 \ {h/{\rm Mpc}}$. Indeed, in that regime, we find that the inferred values of $b_\phi \fnl$ start to substantially deviate from the simulation `truth.' We note that while the cross-power spectrum, $P_{gm}(k)$, is not directly observable, we can access the galaxy-matter projected clustering via joint probes with weak lensing surveys and the CMB map.

\begin{figure}
    \centering    \includegraphics[width=0.48\textwidth]{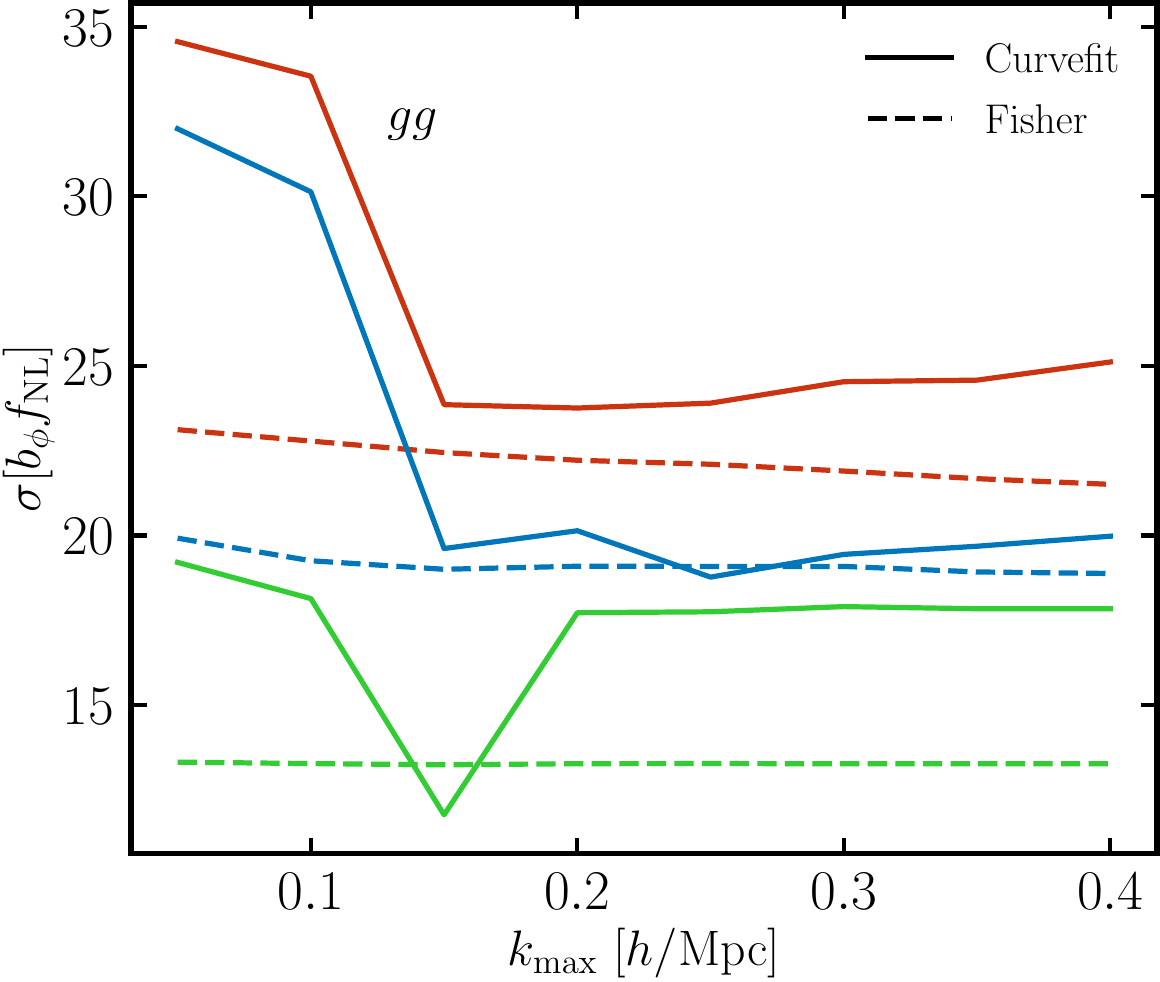}    \includegraphics[width=0.48\textwidth]{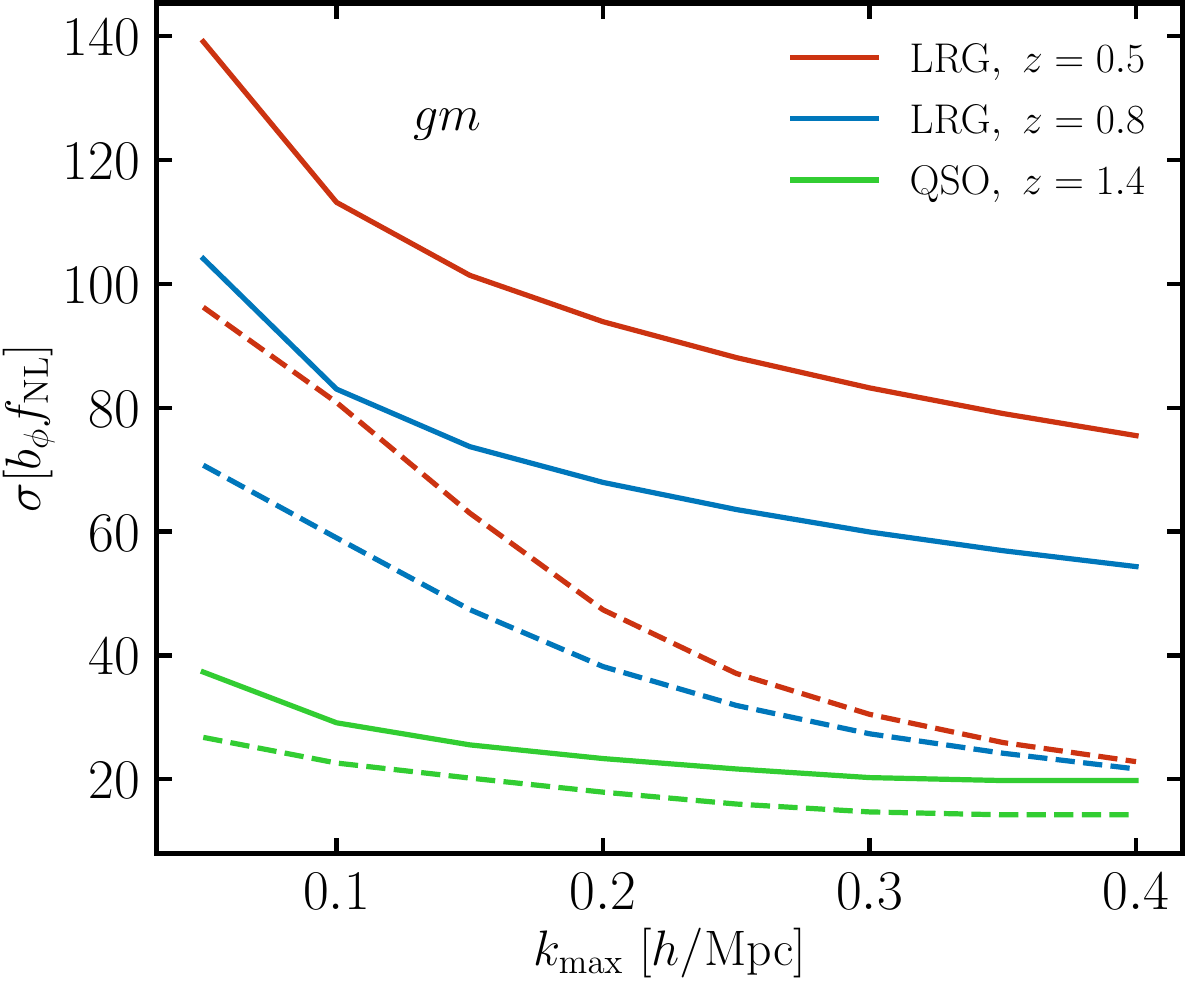}
    \caption{Constraints on the combination $b_\phi \fnl$ as a function of scale ($k_{\rm max}$) for the galaxy auto- ($gg$, upper panel) and cross- ($gm$, lower panel) power spectrum for the LRGs at $z = 0.5$ (red) and $z = 0.8$ (blue) and QSOs at $z = 1.4$ (green). The solid line is obtained via a direct fit (\texttt{curvefit}) to the linear theory model in redshift space, $P_{\ell = 0,2,4}$, whereas the dashed line is obtained via the Fisher approximation (non-marginalized). We see that for the $gg$ case, the agreement between Fisher and the direct fit is very good, though for small values of $k_{\rm max}$, for which there are very few modes available for the fit, there is a larger gap between the two. For the $gm$ case, the agreement is poorer, which we attribute to the approximation adopted for the cross-power spectrum covariance matrix. As we increase $k_{\rm max}$, the constraints on $b_\phi \fnl$ improve only marginally, since the small-scale power spectrum is not affected by the presence of local-type PNG. The constraints for the three tracers from the auto-power spectra are $\sigma[b_\phi \fnl] = 25, \ 20, \ 18$, respectively.
    }
    \label{fig:sigma_bphifnl}
\end{figure}

We present our constraints on $b_\phi \fnl$ in Fig.~\ref{fig:sigma_bphifnl}. The solid line is obtained via a direct fit (using \texttt{curvefit}) to the linear theory model in redshift space, $P_{\ell = 0,2,4}$, whereas the dashed line is obtained via the Fisher approximation (non-marginalized). We see that for the $gg$ case, the agreement between Fisher and the direct fit is very good, though for small values of $k_{\rm max}$, there is a larger gap between the two. We attribute this to the larger noise in the fit, as there are much fewer modes available, and the PNG component is degenerate with the linear bias. For the $gm$ case, the agreement is poorer, which we attribute to the approximation adopted for the cross-power spectrum covariance matrix. 
As we increase $k_{\rm max}$, the constraints on $b_\phi \fnl$ improve only marginally, since the small-scale power spectrum is negligibly affected by the presence of local-type PNG. The constraints for the three tracers from the auto-power spectra are:
\begin{equation}
    \sigma[b_\phi \fnl] = 25, \ 20, \ 18, 
\end{equation}
respectively.

\subsection{Constraints on $\fnl$}
\label{sec:fnl}

\begin{figure}
    \centering    \includegraphics[width=0.48\textwidth]{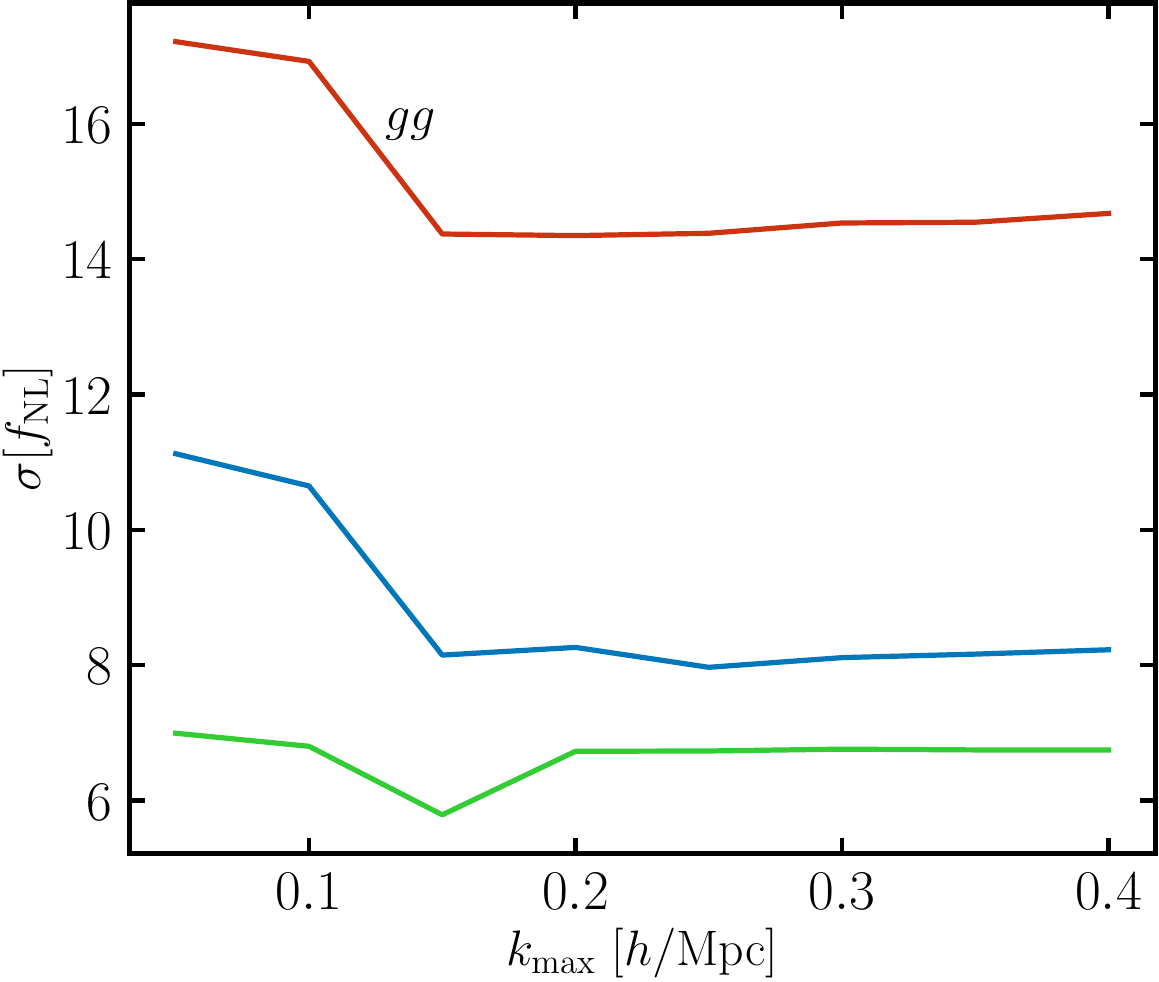}    \includegraphics[width=0.48\textwidth]{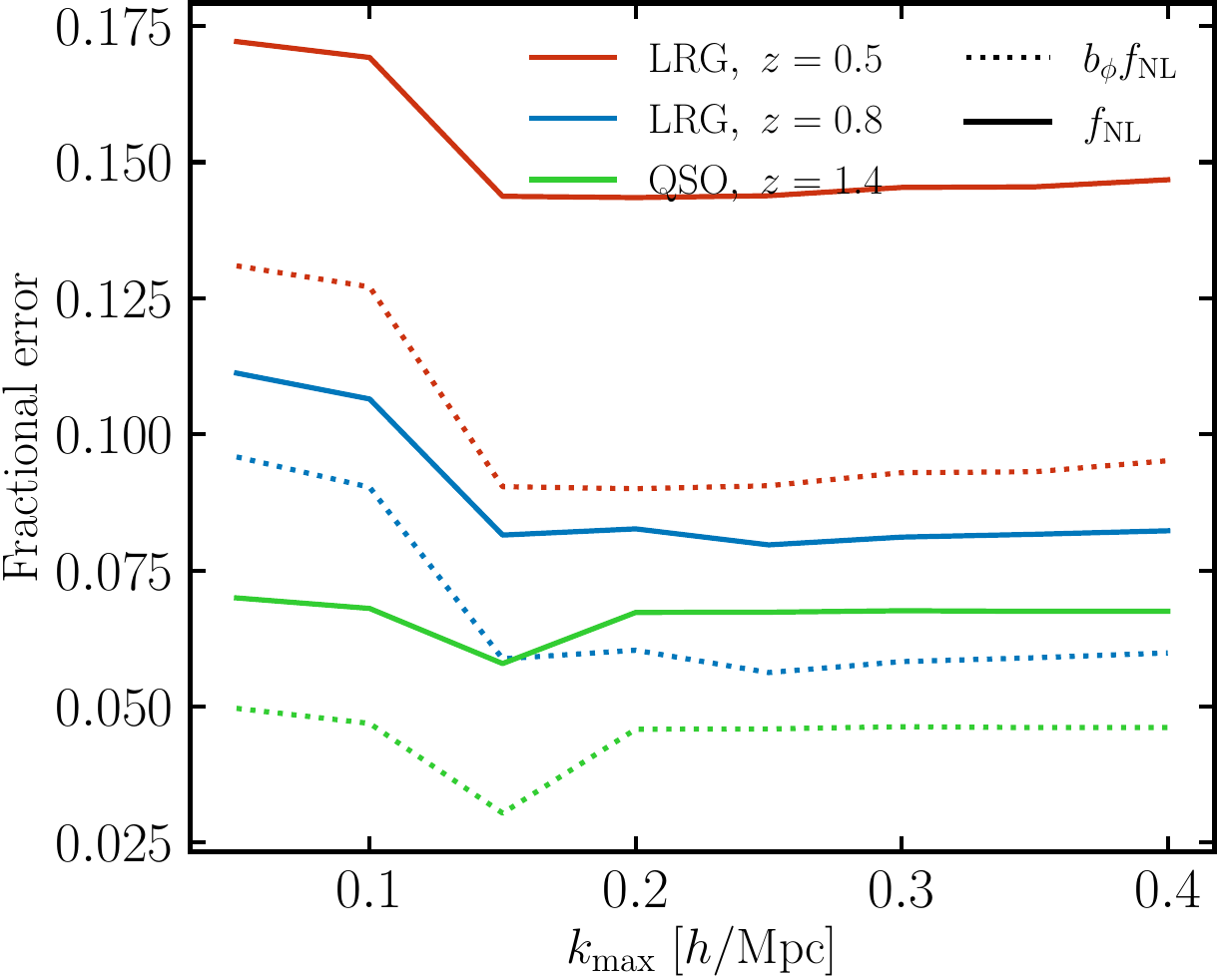}
    \caption{Upper panel: constraints on $\fnl$ as a function of scale ($k_{\rm max}$) for the galaxy auto-power spectrum of the LRGs at $z = 0.5$ (red) and $z = 0.8$ (blue) and QSOs at $z = 1.4$ (green). We see that the best constraints are obtained from the LRGs at $z = 0.8$, which have a larger volume. While the QSOs had poorer constraints on $b_\phi \fnl$ compared with the other two tracers, we see that they yield comparable $\sigma[\fnl]$ to the LRGs at $z = 0.5$, as $b_\phi$ is highest for that sample. 
    We quote the minimum error on $\fnl$ after taking into account the uncertainty of $b_\phi$ for each of the three samples as $\sigma[\fnl] = 15, \ 8, \ 7$, respectively. 
Lower panel: Fractional error, defined as $\sigma[\cdot]/[\cdot]$, of the combination $b_\phi \fnl$ (dotted line) and just $\fnl$ using Eq.~\ref{eq:error} and the cosntraints on $b_\phi$ from Section~\ref{sec:bphi}. For the LRGs at $z = 0.8$ and the QSOs, the fractional error on $\fnl$ compared with that on $b_\phi \fnl$ worsens by about a factor of 1.5 due to the large uncertainty on the astrophysical parameter, $b_\phi$, and by 2 for the LRGs at $z = 0.5$. This shows that it is of great importance to understand the galaxy-halo connection of the DESI samples in order to reduce both the uncertainty and the bias on the inferred value of $\fnl$.} 
    \label{fig:sigma_fnl}
\end{figure}

Taking into account our findings from Section~\ref{sec:bphi} on the uncertainty of $b_\phi$, we can now translate the uncertainty on the combination $b_\phi \fnl$ into an uncertainty on $\fnl$ \BH{using linear error propagation}, as follows:
\begin{equation}
    \sigma[\fnl] = \fnl \left(\left(\frac{\sigma[b_\phi]}{b_\phi}\right)^2 + \left(\frac{\sigma[b_\phi \fnl]}{b_\phi \fnl}\right)^2 - 2\frac{\sigma[\BH{b_\phi, \fnl}]}{b_\phi \fnl}\right)^{1/2} .
\label{eq:error}
\end{equation}
Assuming that $b_\phi$ and $\fnl$ have negligible covariance, we can simplify the equation above by ignoring the cross-term, $\sigma[\BH{b_\phi,\fnl}] \approx 0$. Substituting the standard deviation and fiducial values of $b_\phi$ for the three samples from Section~\ref{sec:bphi} and the $\sigma[b_\phi \fnl]$ from the Fisher analysis, we obtain the following constraints on $\fnl$ from the galaxy auto-power spectrum:
\begin{equation}
    \sigma[\fnl] = 15, \ 8, \ 7. 
\end{equation}
\BH{In this analysis, we assume that the other cosmological parameters are kept fixed.}
These constraints are shown in Fig.~\ref{fig:sigma_fnl}, where we see the constraints as a function of scale ($k_{\rm max}$). 
The best constraints are obtained from the QSOs at $z = 1.4$, as expected, as quasars have the largest volume and largest value of $b_\phi$. The LRGs at $z = 0.8$ yield very comparable constraints, as their $b_\phi$ is relatively high and well constrained. 
In the figure, we also display the fractional error, $\sigma[\cdot]/[\cdot]$, of the combination $b_\phi \fnl$ and just $\fnl$ using Eq.~\ref{eq:error} and the constraints on $b_\phi$ from Section~\ref{sec:bphi}. For the LRGs at $z = 0.8$ and the QSOs, the fractional error on $\fnl$ compared with that on $b_\phi \fnl$ is a factor of 1.5 worse due to the large uncertainty on the astrophysical parameter, $b_\phi$. For the LRGs at $z = 0.5$, the two curves differ by almost a factor of 2, as $b_\phi$ is poorly constrained for that sample (see Fig.~\ref{fig:bphi_gal}). This finding hints that more investigation into the complex link between galaxy formation and local-type PNG is warranted.
In order to place $\sigma[\fnl] \sim 1$ constraints on this model of the primordial Universe, we need to not only make precise measurements on large-scales using spectroscopic or weak lensing surveys of large volumes, but also substantially improve our constraints on the galaxy-halo connection and thus, $b_\phi$.

\section{Summary and conclusions}
\label{sec:conc}

Observations of the CMB have revolutionized our understanding of the primordial Universe, demonstrating that primordial fluctuations were nearly homogeneous and isotropic with an almost scale-invariant power spectrum. To explain these observations, a number of early Universe theories such as inflationary and ekpyrotic models have been proposed. In order to discriminate between these models, we need to attain definitive evidence of their predicted byproducts such as signatures of primordial gravitational waves in the $B$-mode polarization of the CMB and primordial non-Gaussianity (PNG) \citep{2010AdAst2010E..67L,2016ASSP...45...41M,2019BAAS...51c.107M}. PNG is an invaluable probe, as it can reveal information about the field content, dynamics, and strength of interactions in the early Universe. While measurements of the CMB have placed tight constraints on the presence of PNG, large-scale structure probes are potentially even more powerful, as they can access a larger number of Fourier modes corresponding to the volume covered by the survey.

In the next few years, large-scale structure experiments such as DESI and \textit{Roman} will catalogue the 3D positions of tens of millions of galaxies, mapping an unprecedentedly large volume of space. Extracting sub-percent comparisons between survey observations and cosmological predictions of PNG requires high-precision mock data and robust theoretical models. Although approximate methods are capable of generating sample data, cosmological $N$-body simulations are a central tool for making accurate forecasts for the non-linear regime of gravitational structure formation. Moreover, extracting PNG constraints from large-scale structure is challenging due to complex observational effects (such as survey depth, mask, fiber collisions), modeling of the summary statistics, and disentangling astrophysical from primordial physics effects. Understanding the relation between PNG and late-time observables is the main objective of the \textsc{AbacusPNG} set of $N$-body simulations, presented in this work.

Here, we present a new set of simulations, \textsc{AbacusPNG}, which was run with the extremely fast and accurate \texttt{Abacus} code \citep{2021MNRAS.508..575G}, modified to possess a non-Gaussian primordial gravitational potential. The aim of this work is to aid current and next generation of surveys in constraining local-type PNG from large-scale structure and to validate the simulation methods for future larger simulation suites with \texttt{Abacus}.  
First, we have introduced the \textsc{AbacusPNG} set in Section~\ref{sec:sims}. Next, we have validated its initial conditions by studying the matter bispectrum and comparing it against the tree-level prediction (see Fig.~\ref{fig:bk_matter}). In addition, we have compared the matter power spectrum at late times with the one-loop theoretical prediction, finding good agreement on the relevant scales (see Fig.~\ref{fig:pk_matter}). Finally, we have tested that the $b_\phi$ we compute from the PNG simulations is in agreement with the value of the parameter inferred using the `separate universe' technique on the original linear-derivative \textsc{AbacusSummit} boxes (see Fig.~\ref{fig:bphi}). We also quote, the slope of the universality relation for the Abacus catalogs as $c \approx 0.85$ (see Eq.~\ref{eq:bphiuniv}).

As a science case, we focus on the halo and galaxy fields in Section~\ref{sec:halos} and Section~\ref{sec:desi}, respectively. First, we study the halo response to different assembly bias properties, which have been associated with the galaxy samples of modern surveys such as BOSS and DESI \citep[see][for a review]{2018ARA&A..56..435W}. We find (see Fig.~\ref{fig:bphi_hab}) a strong response to all three parameters we study: concentration, shear and accretion rate, suggesting that there are additional non-negligible dependencies of $b_\phi$ on galaxy and halo properties besides halo mass (linear bias). 

In Section~\ref{sec:bphi}, we present the first study of the parameter $b_\phi$ estimated from realistic galaxy samples obtained from fits to the DESI data \citep{2023arXiv230606314Y}. In particular, focussing on the two most relevant tracers, LRGs and QSOs, we study the value of $b_\phi$ for slight variations of the main and extended (i.e., pertaining to assembly bias) HOD parameters that are still allowed by the data (see Fig.~\ref{fig:bphi_gal}). We find that the uncertainty on $b_\phi$ is 11\%, 6\%, 5\% for LRGs at $z = 0.5$ and $z = 0.8$ 
and QSOs at $z = 1.4$, respectively. We then translate these values of $b_\phi$ into constraints on $\fnl$ from a simple Fisher analysis using linear theory. We find that the inferred error of $\fnl$, once we take into account the uncertainty in $b_\phi$, becomes 15, 8, 7 for the three tracers, 
respectively (see Fig.~\ref{fig:sigma_fnl}), which is about a factor of 1.5-2 worse compared with the case in which $b_\phi$ is known to a high precision. This suggests that understanding the response of galaxy formation to local-type PNG as well as the assembly bias properties of a given galaxy tracer is essential if we wish to perform unbiased, high-precision measurements of local-type PNG. For future work, we leave the exploration of the connection between observable galaxy properties and $b_\phi$ via hydrodynamical and $N$-body simulations, which would allow us to construct multi-tracer samples and thus greatly enhance our constraints on $\fnl$.

While we are a long way away from uncovering the subtle interplay between the physics of the primordial Universe and the observed large-scale properties of the Universe, our hope is that the \textsc{AbacusPNG} simulations presented in this study will aid us in this journey and provide us with some of the missing pieces needed to unravel the elusive link between fundamental cosmology and astrophysics.

\begin{acknowledgments}
We thank Dionysis Karagiannis, Will Coulton, Xinyi Chen, Adrian Gutierrez, Jamie Sullivan, Adrian Bayer, and Sihan (Sandy) Yuan for illuminating discussions during the preparation of this draft.
BH is supported by the Miller Institute for Basic Science. DJE is supported by the U.S. Department of Energy grant DE-SC0007881 and as a Simons Foundation Investigator.
SF is supported by Lawrence Berkeley National Laboratory and the Director, Office of Science, Office of High Energy Physics of the U.S. Department of Energy under Contract No.\ DE-AC02-05CH11231.

This research used resources of the National Energy Research Scientific Computing Center, which is supported by the Office of Science of the U.S. Department of Energy under Contract No.~DE-AC02-05CH11231.  Additional computations in this work were performed at facilities supported by the Scientific Computing Core at the Flatiron Institute, a division of the Simons Foundation.

\end{acknowledgments}

\section*{Data Availability}
The simulations used in this work are publicly available. Instructions for access and download are given at \url{https://abacussummit.readthedocs.io/en/latest/data-access.html}.

\bibliography{refs}{}
\bibliographystyle{prsty}



\end{document}